\documentclass[aps,manuscript,showpacs,showkeys,superscriptaddress]{revtex4}
\usepackage{graphicx}
\usepackage{epsfig}  
\usepackage{epsf}    
\usepackage{dcolumn}
\usepackage{bm}
\usepackage{dcolumn}
\usepackage{textcomp}
\usepackage[tbtags]{amsmath}
\usepackage{amsfonts}
\usepackage{float}
\usepackage{subfig}
\usepackage[]{hyperref}
  \hypersetup{
  unicode=false,          
  pdftoolbar=true,        
  pdfmenubar=true,        
  pdffitwindow=true,     
  pdfstartview={FitH},    
  pdfsubject={Getting the best out of T2K and NOvA},   
  pdfnewwindow=true,      
  pdfcreator={RevTeX},
  colorlinks=true,       
  linkcolor=red,          
  citecolor=blue,        
  urlcolor=blue,           
  }
\usepackage{hypcap}

\newcommand{\beq}{\begin{equation}}
\newcommand{\eeq}{\end{equation}}
\newcommand{\beqa}{\begin{eqnarray}}
\newcommand{\eeqa}{\end{eqnarray}}

\newcommand{\tx}{{\theta_{12}}}
\newcommand{\ty}{{\theta_{13}}}
\newcommand{\tz}{{\theta_{23}}}

\newcommand{\dl}{{\Delta_{31}}}
\newcommand{\ds}{{\Delta_{21}}}

\newcommand{\dcp}{{\delta_{CP}}}
\newcommand{\nova}{{NO$\nu$A }}
\newcommand{\pmue}{P_{\mu e}}
\newcommand{\pmuebar}{P_{\bar{\mu} \bar{e}}}

\begin{document}

\preprint{TIFR/TH/12-03}

\title{Getting the best out of T2K and \nova}


\author{Suprabh Prakash}
\email[Email Address: ]{suprabh@phy.iitb.ac.in}
\affiliation{
Department of Physics, Indian Institute of Technology Bombay,
Mumbai 400076, India}

\author{Sushant K. Raut}
\email[Email Address: ]{sushant@phy.iitb.ac.in}
\affiliation{
Department of Physics, Indian Institute of Technology Bombay,
Mumbai 400076, India}
 
\author{S. Uma Sankar \footnote{Corresponding author}}
\email[Email Address: ]{uma@phy.iitb.ac.in}
\affiliation{
Department of Physics, Indian Institute of Technology Bombay,
Mumbai 400076, India}
\affiliation{
Department of Theoretical Physics, Tata Institute of Fundamental Research,
Mumbai 400005, India}
\date{\today}
\begin{abstract}
We explore the combined physics potential of T2K and \nova in light
of the moderately large measured value of $\ty$.
For $\sin^2 2 \theta_{13} = 0.1$, which is close to the 
best fit value, a 90 \% C.L. evidence for the hierarchy can be 
obtained only for the combinations 
(Normal hierarchy, $-170^\circ \leq \dcp \leq 0^\circ$) and (Inverted 
hierarchy, $0^\circ \leq \dcp \leq 170^\circ$),
with the currently planned runs of
\nova and T2K. However, the hierarchy can essentially 
be determined for any value of $\dcp$,
if the statistics of \nova are increased by $50\%$
and those of T2K are doubled. 
Such an increase will also give an allowed region of $\dcp$
around the its true value, except for the CP conserving
cases $\dcp = 0 \ {\rm or} \pm 180^\circ$. 
We demonstrate that 
any measurement of $\dcp$ is not possible without
first determining the hierarchy. 
We find that comparable data from a shorter baseline
(L $\sim$ 130 km) experiment will not lead to any significant
improvement.
\end{abstract}
\maketitle

\section{Introduction}
Neutrino physics has entered a phase of precision measurements.
During the past few years, the following precise measurements 
of neutrino parameters have
been made with high intensity sources: 
\begin{itemize}
\item
The smaller mass-squared difference $\ds = m_2^2 - m_1^2$ 
is measured by KamLAND \cite{kamland} while the precision on $\tx$ is  
controlled by the solar experiments \cite{sno}. Global analysis
of all the data, in the three flavour oscillation framework, gives
$\ds = (7.6 \pm 0.2) \times 10^{-5}$ eV$^2$ and $\sin^2 \tx 
= 0.312 \pm 0.016$ \cite{globalfit_2011}.
\item
MINOS \cite{minos} experiment has measured the magnitude of the
mass-squared difference in the $\nu_\mu$ survival probability. 
The precision on $\tz$ is controlled by atmospheric neutrino 
data \cite{PRD81.092004}. Global analysis
gives two distinct values of $\dl$ depending on whether it
is positive [which is the case for normal hierarchy (NH)] or
negative [which is the case for inverted hierarchy (IH)]. The
ranges are $\dl(NH) = (2.45 \pm 0.09) \times 10^{-3}$ eV$^2$
and $\dl(IH) = (-2.31 \pm 0.09) \times 10^{-3}$ eV$^2$ with
$\sin^2 \tz = 0.51 \pm 0.06$ for both cases \cite{globalfit_2011}. 
\item
The global fits to data from the accelerator experiments T2K 
\cite{t2k_t13_jun2011} and MINOS \cite{minos_t13_jun2011}
and the reactor experiments DChooz \cite{dchooz_dec2011}, 
Daya Bay \cite{dayabay2012} and RENO \cite{reno2012} have determined $\ty$ 
to be non-zero at $5 \sigma$ level, with the best fit very close to 
$\sin^2 2 \theta_{13} \simeq 0.1$ \cite{globalfit_t13,fogli2012}. 
\end{itemize}

We expect the following improvements in precision during the
next few years.

\begin{itemize}
\item
Very high statistics data from  T2K \cite{t2k} and MINOS \cite{minos} 
experiments will 
improve the precision on $|\dl|$ and $\sin^2 2 \tz$ to a few 
percent level.
\item
Reactor experiments are taking further data 
\cite{dchooz1,dchooz2,dayabay,reno}.
The survival probability at these reactor experiments  
is sensitive only to the mixing angle $\ty$ and hence they can 
measure this angle unambiguously. By the time they finish 
running (around 2016), we estimate that they should be able to 
measure $\sin^2 2 \ty$ to a precision of about $0.005$. 
\end{itemize}

In light of these current and expected near future measurements,
the next goals of neutrino oscillation experiments are the 
determination of neutrino mass hierarchy, detection of CP
violation in the leptonic sector and measurement of $\dcp$. 
These goals can be achieved by high statistics
accelerator experiments measuring $\nu_\mu \to \nu_e$ and
$\bar{\nu}_\mu \to \bar{\nu}_e$ oscillation probabilities.
Among such experiments, T2K is presently taking data and 
\nova is under construction and is expected to start 
taking data around 2014. All other experiments, capable of making
these measurements, are far off in future. In this paper,
we study the combined ability of T2K and \nova to achieve
the above goals.

In the above discussion, we have two different magnitudes for $\dl$
for the two hierarchies because the mass-squared difference
measured in $\nu_\mu$ survival probability is not $\dl$ but is an
effective one defined by \cite{parke_defn,degouvea_defn}
\beq
\Delta m^2_{\mu \mu} = \dl - \left(\cos^2\tx - \cos \dcp \sin\ty \sin 2\tx \tan \tz\right)\ds.
\eeq
Accelerator experiments, such as MINOS and T2K, measure the 
magnitude of the above quantity. But the magnitudes of $\dl$
will turn out to be different for $\dl$ positive (NH) and 
$\dl$ negative (IH).

\section{Simulation Details}

Before discussing various physics issues, we discuss the 
details of our simulation. We do this because we will 
illustrate various points through the means of simulation.

We use the software GLoBES \cite{globes1,globes2} for simulating the data of 
T2K, \nova and an envisaged short baseline experiment from CERN to
Fr\'{e}jus (C2F), which is a scaled down version of MEMPHYS \cite{t2k,twobase1,globes_t2k3,nova,globes_nova,campagne,
globes_spl2,globes_spl3,messier_xsec,paschos_xsec,globes_t2k4,globes_t2k5}. 
Various details of these experiments and their
characteristics, especially the signal and background
acceptances, are given in Table~\ref{tab1}. The basic properties of 
\nova are taken from Ref.~\cite{nova} and of T2K are taken from Ref.~\cite{t2k}.
The efficiencies for each of the experiments are taken from GLoBES
\cite{globes1,globes2}. The background errors consist of errors in
flux normalization (norm) and in spectrum (tilt).

\begin{table}
	\begin{tabular}{  l  |p{3.8cm} | p{3.8cm} | p{3.8cm} }
	
         \hline
         \hline
	Characteristic &NO$\nu$A&T2K&C2F (assumed) \\
	\hline
         \hline
	 Baseline & 812 km & 295 km & 130 km \\
	
	 Location & Fermilab - Ash River& J-PARC - Kamioka& CERN - Fr\'{e}jus \\
	
	 Beam&NuMI beam 0.8$^\circ$ off - axis&JHF beam 2.5$^\circ$ off - axis&SPL superbeam  \\
	
	 Beam power&0.7 MW&0.75 MW&0.75 MW  \\

	 Flux peaks at&2 GeV&0.6 GeV&0.35 GeV  \\
	
	 $\pmue$ 1st Osc. Maximum&1.5 GeV&0.55 GeV&0.25 GeV \\
	
	 Detector&TASD, 15 kton&Water \v{C}erenkov, 22.5 kton&Water \v{C}erenkov, 22.5 kton \\ 

          Runtime (years)&3 in $\nu$ + 3 in $\bar{\nu}$&5 in $\nu$&3 in $\nu$ + 3 in $\bar{\nu}$  \\
          
          Signal 1 (acceptance)&$\nu_{e}$ appearance(26$\%$)&$\nu_{e}$ appearance(87$\%$)& $\nu_{e}$ appearance(71$\%$)\\

          Signal 1 error&5$\%$, 2.5$\%$&2$\%$, 1$\%$&2$\%$, 0.01$\%$  \\
          (norm.,tilt)&&&\\

          Background 1 (acceptance)&mis - id muons/anti - muons(0.13$\%$), NC events(0.28$\%$), Beam $\nu_{e}$/$\bar{\nu}_{e}$(16$\%$)
                      &mis - id muons/anti - muons, NC events, Beam $\nu_{e}$/$\bar{\nu}_{e}$(binned events from 
                       GLoBES \cite{globes1,globes2})
                      &mis - id muons/anti - muons(0.054$\%$), NC events(0.065$\%$), Beam $\nu_{e}$/$\bar{\nu}_{e}$(70$\%$)  \\

          Background 1 error&10$\%$, 2.5$\%$&20$\%$, 5$\%$&2$\%$, 0.01$\%$  \\
          (norm.,tilt)&&&\\

          Signal 2 (acceptance)&$\bar{\nu}_{e}$ appearance(41$\%$)&$\bar{\nu}_{e}$ appearance(87$\%$)& $\bar{\nu}_{e}$ appearance(68$\%$)  \\

          Signal 2 error&5$\%$, 2.5$\%$&2$\%$, 1$\%$&2$\%$, 0.01$\%$  \\
          (norm.,tilt)&&&\\

          Background 2 (acceptance)&mis - id muons/anti - muons(0.13$\%$), NC events(0.88$\%$), Beam $\nu_{e}$/$\bar{\nu}_{e}$(33.6$\%$)
                      &mis - id muons/anti - muons, NC events, Beam $\nu_{e}$/$\bar{\nu}_{e}$(binned events from 
                       GLoBES \cite{globes1,globes2})
                      &mis - id muons/anti - muons(0.054$\%$), NC events(0.25$\%$), Beam $\nu_{e}$/$\bar{\nu}_{e}$(70$\%$)  \\

           Background 2 error&10$\%$, 2.5$\%$&20$\%$, 5$\%$&2$\%$, 0.01$\%$  \\
          (norm.,tilt)&&&\\
        \hline
        \hline
	\end{tabular}
\caption{Properties of various long baseline experiments considered 
in this paper.} 
\label{tab1}
\end{table} 

\vspace{1cm}

We have kept the solar parameters $\ds$ and $\tx$ fixed at their
best fit values throughout the calculation. We have taken the central 
values of $|\dl|$ and $\tz$ to be their best fit values. 
We took  
$\sigma\left(\sin^22\tz\right) = 0.02$ and 
$\sigma\left(|\dl|\right) = 0.03\times\left(|\dl|\right)$,
because of the precision expected from T2K.
We have done computations for various different values of
$\sin^2 2 \ty$ in the range $0.05 - 0.2$ \cite{globalfit_t13,fogli2012}. 
We took  $\sigma\left(\sin^22\ty\right) = 0.005$ which is
the final precision we can hope for from the reactor experiments.
The value
of the CP-violating phase $\dcp$ is varied over its 
entire range $-180^\circ$ to $180^\circ$.

We compute statistical $\chi_{st}^2$ as
\beq
\chi_{st}^2 = \sum_i \frac{(N_i^{true} - N_i^{test})^2}{N_i^{true}}, 
\eeq
where $N_i^{true}$ is the event distribution for true hierarchy
and some fixed true value of $\dcp$. $N_i^{test}$ is the event 
distribution with the test hierarchy either true or wrong 
and a varying test value of $\dcp$
as inputs. The index $i$ runs over the number of energy bins. 
The final $\chi^2$ is computed including the systematic errors, 
described in Table~\ref{tab1}, and the priors on $|\dl|$, $\sin^2 2 \tz$ 
and $\sin^2 2 \ty$.
The prior on $\sin^22\ty$ effectively takes into account the data 
due to reactor neutrino experiments.

In the following we consider two kinds of plots both of which
are shown as contours in the $\sin^2 2 \ty$-$\dcp$ plane.
\begin{itemize}
\item Hierarchy exclusion plots: These are plotted in 
the plane of true values of $\sin^2 2 \ty$-$\dcp$. 
The contours in these plots define the line $\chi^2 = 2.71$.
In computing this $\chi^2$, we have marginalized over 
the parameter ranges described above.
For all sets of parameter values to the right of the contour, the
wrong hierarchy can be ruled out at $90 \%$ C.L. 
\item Allowed region plots: These are plotted in the plane of
test values of $\sin^2 2 \ty$-$\dcp$. The contours in these plots 
are defined by $\chi^2 = 4.61$. The region enclosed by them
is the set of allowed values of $\sin^2 2 \ty$-$\dcp$
at $90 \%$ C.L. for a given set of neutrino parameters.
\end{itemize}

Throughout this paper, the phrase "hierarchy determination"
implies 90$\%$ C.L. evidence for hierarchy.

\section{Hierarchy determination with $\pmue$}

The $\nu_{\mu} \to \nu_{e}$ channel is sensitive to a number of
neutrino parameters and hence is the most sought after in the 
study of neutrino oscillation physics using long baseline experiments.
In the presence of matter, the $\nu_{\mu} \to \nu_{e}$ oscillation 
probability, expanded perturbatively in the small mass-squared 
difference, $\ds$ is given by \cite{cervera,akhmedov,freund} 
\beqa
\nonumber P\left(\nu_{\mu}\to \nu_{e}\right) &=& P_{\mu e} = 
\sin^2 2\ty\sin^2\tz
\frac{\sin^2\hat{\Delta}(1 - \hat{A})}{(1 - \hat{A})^2}  \nonumber \\ 
&& 
+\alpha\cos\ty\sin2\tx\sin2\ty\sin2\tz\cos(\hat{\Delta}+\dcp)\frac{\sin\hat{\Delta}\hat{A}}{\hat{A}}
\frac{\sin\hat{\Delta}(1-\hat{A})}{1-\hat{A}} \nonumber \\
&&
 + \alpha^2\sin^22\tx\cos^2\ty\cos^2\tz\frac{\sin^2\hat{\Delta}\hat{A}}{\hat{A}^2}  
\label{pmue}
\eeqa
where $\hat{\Delta}= \dl L/4E$, $\hat{A}= A/\dl$, $\alpha = \ds/\dl$.
$A$ is the Wolfenstein matter term \cite{msw1} and is given by
$A ({\rm eV}^2) = 0.76 \times 10^{-4} \rho \ ({\rm gm/cc}) E ({\rm GeV})$.

For NH $\dl$ is positive and for IH
$\dl$ is negative. The matter term $A$ is positive for neutrinos
and is negative for anti-neutrinos. Hence, in neutrino oscillation
probability, $\hat{A}$ is positive for NH and is negative for IH.
For anti-neutrinos, $\hat{A}$ is negative
for NH and positive for IH and the sign of $\dcp$ is
reversed. The presence of the term $\hat{A}$ in $P_{\mu e}$
and in $P_{\bar{\mu} \bar{e}}$ makes them sensitive to hierarchy.
The longer the baseline of an experiment, the greater is the 
sensitivity to hierarchy because, $P_{\mu e}$ peaks at a higher
energy for longer baseline and the matter term is larger for 
higher energies.  
 
As can be seen from Eq.~(\ref{pmue}), 
$\pmue$ is dependent on $\ty$, hierarchy amd $\dcp$
in addition to other well determined parameters. A measurement of 
this quantity will not give us a unique solution of neutrino
parameters but instead will lead to a number of degenerate 
solutions \cite{degeneracy1,degeneracy2,degeneracy3,degeneracy4}. 
Since $\ty$ is measured unambiguously and precisely 
\cite{dchooz_dec2011,dayabay2012,reno2012}, degeneracies involving this 
parameter are no longer relevant. Only hierarchy-$\dcp$ 
degeneracy has to be considered. This degeneracy prevents any one experiment
from determining hierarchy and $\dcp$, leading to the need for
data from two or more long baseline experiments \cite{twobase1,twobase2,twobase3,twobase4,twobase5}. 

\subsection{Hierarchy-$\dcp$ degeneracy for \nova}

\begin{figure}[H]
\vspace{-1.7cm}
\begin{center}
	\epsfig{file=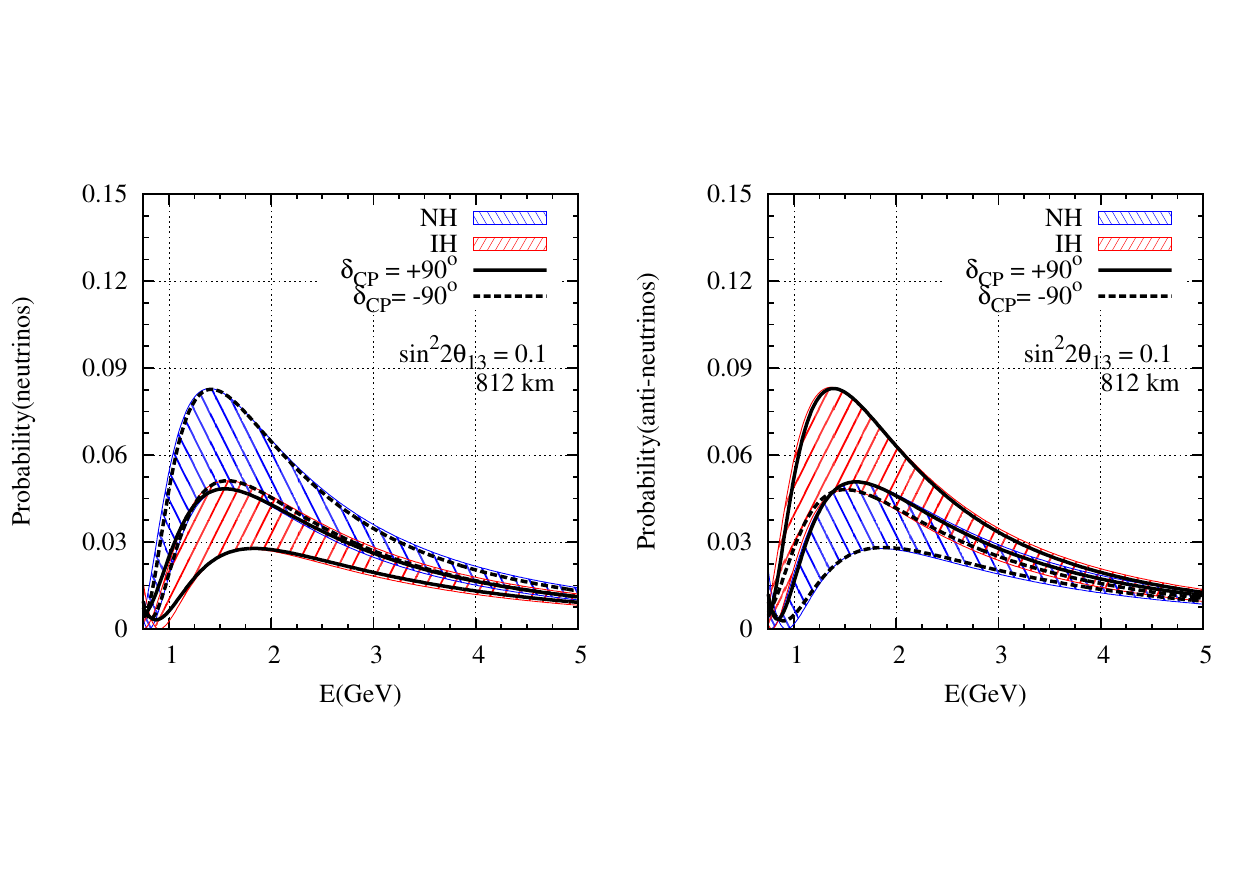,width=1\textwidth}
\end{center}
\vspace{-2.7cm}
\caption{\label{pmuenova}\footnotesize
(colour online) $\pmue$ (left panel) and $\pmuebar$ (right panel) 
bands for NO$\nu$A for $\sin^2 2\ty = 0.1$ }
\end{figure}

First we consider the hierarchy determination capacity of \nova alone 
because the matter term and the hierarchy dependence is the largest
for this experiment, due to the flux peaking at higher energy. 
In Fig.~\ref{pmuenova} (left panel), we have plotted $\pmue$ vs E for
both NH and for IH for \nova baseline of 812 km. 
The bands correspond to the variation of $\dcp$ from $-180^\circ$ to 
$+180^\circ$.  
The values of $\pmue$ are, in general, higher for 
NH and lower for IH. This is a straightforward consequence of
the $\hat{A}$ dependence of $\pmue$. Further, we note that 
for both NH and IH, the value of $\dcp = +90^\circ$ gives the 
lowest curve in the band and the value of $\dcp = -90^\circ$ gives 
the highest curve in the band. This behaviour can also be 
easily understood from Eq.~(\ref{pmue}). At the oscillation
maximum, $\hat{\Delta} \simeq 90^\circ$. Hence $\cos(\hat{\Delta}
+\dcp)$ is $+1$ for $\dcp = -90^\circ$ and is $-1$ for $\dcp = 
+90^\circ$. As can be seen from the figure, there is an overlap 
of the bands for (NH, $\dcp \simeq +90^\circ$) and 
(IH, $\dcp \simeq -90^\circ$).
Hence, if the measured probability comes to be these values,
then we have two degenerate solutions. In Fig.~\ref{pmuenova} (right panel), 
we have plotted the corresponding anti-neutrino probabilities.
$\pmuebar$ is higher for IH and lower for NH as a consequence of
the reversal of $\hat{A}$ sign. Since $\dcp$ sign is reversed
for anti-neutrinos, here 
$\dcp = + 90^\circ$ defines the upper curves and $\dcp = 
-90^\circ$ defines the lower curves. Here again there is an
overlap between (NH, $\dcp \simeq +90^\circ$) and (IH, $\dcp \simeq -90^\circ$)
so we get the same degenerate solutions as the neutrino case.  

From Fig.~\ref{pmuenova},
we can define the concept of favourable half plane for
each hierarchy. Suppose NH is the true hierarchy. If 
$\dcp$ is in the lower half plane ($-180^\circ \leq \dcp \leq 0^\circ$, LHP)
then all the curves for $\pmue(NH,\dcp)$ lie much above
the set of curves for $\pmue(IH,\dcp)$. In the case of
anti-neutrinos, $\pmuebar(NH, \dcp)$ will be much below 
$\pmuebar(IH, \dcp)$. In such a situation, the data from
\nova alone can determine the hierarchy. Therefore we call
the LHP to be the favourable half-plane for NH. Similar
arguments hold if IH is true hierarchy and $\dcp$ is in
the upper half plane (UHP). So UHP is the favourable
half plane for IH. Thus, if nature chooses one of the
following two combinations (NH, LHP) or (IH, UHP), then
\nova, by itself, can determine the hierarchy.

\begin{figure}[H]
\vspace{-1.7cm}
\begin{center}
	\epsfig{file=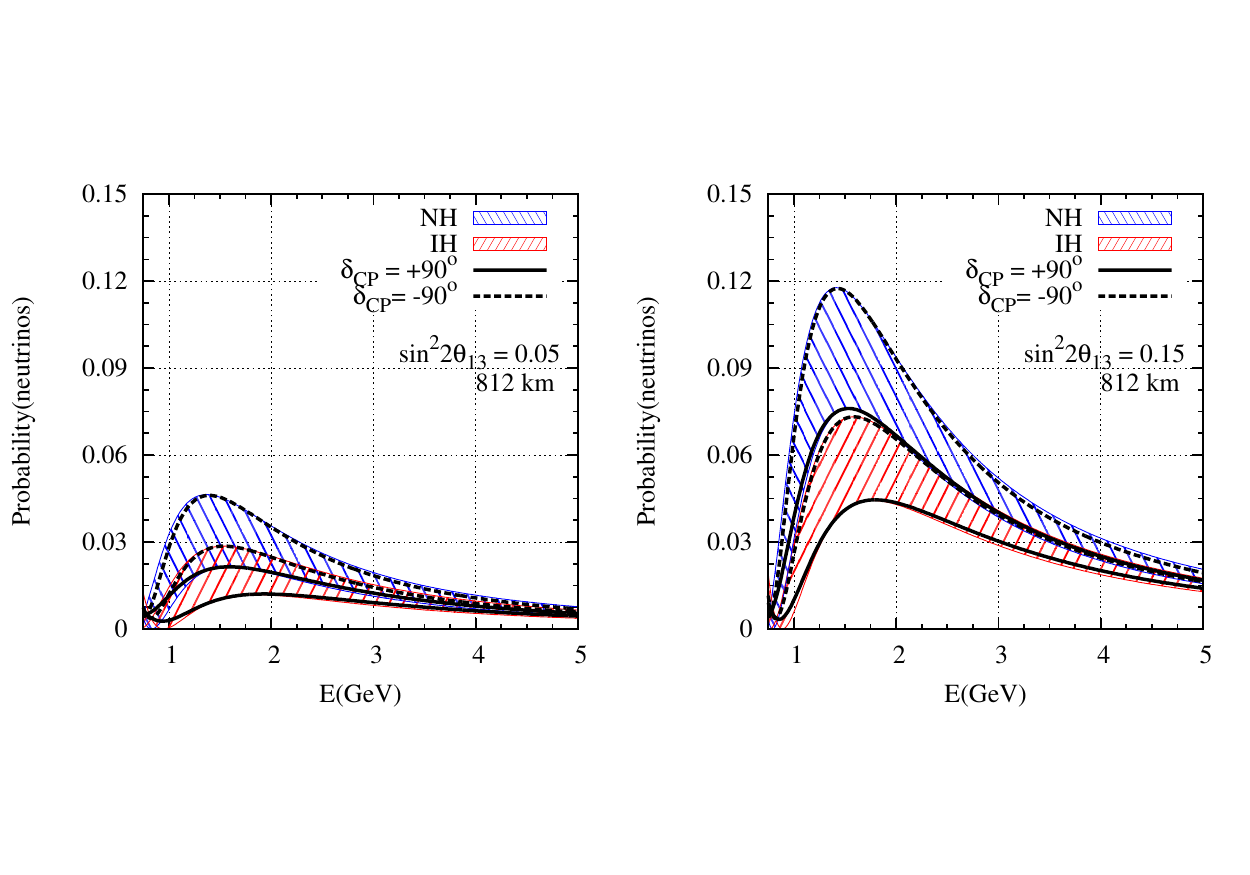,width=1\textwidth}
\end{center}
\vspace{-2.7cm}
\caption{\label{pmuet13}\footnotesize
(colour online) $\pmue$ bands for NO$\nu$A for  
$\sin^2 2\ty = 0.05$ (left panel) and $0.15$ (right panel)}
\end{figure}

The separation between the set of curves 
$\pmue(NH,\dcp)$ and $\pmue(IH,\dcp)$ also depends on
$\ty$. The two sets have more overlap for smaller
values of $\ty$ but become more separated for larger
values of $\ty$. This is illustrated in
Fig.~\ref{pmuet13}, showing $\pmue$ vs E, for a lower and higher
value of $\sin^2 2 \ty$. It is easier to determine the 
hierarchy if the separation between the curves is larger,
that is if $\ty$ is larger. 

The favourable and unfavourable half planes for a 
particular hierarchy can also be defined from Eq.~(\ref{pmue}),
where the $\dcp$ dependence occurs purely in the form
$\cos(\hat{\Delta} + \dcp)$. If NH is the true hierarchy, $\hat{\Delta} \approx 90^\circ$
around the probability maximum.
Then, the $\dcp$ dependent term increases $\pmue$
if $\dcp$ is in the LHP and decreases it if $\dcp$ is in the UHP.
Hence a cleaner separation from $\pmue(IH,\dcp)$ can be
obtained only if $\dcp$ is in the LHP. If IH is the true hierarchy,
$\hat{\Delta} \approx -90^\circ$. Then $\pmue$ is reduced,
and moved away from $\pmue(NH,\dcp)$ if $\dcp$ is in the
UHP. Thus UHP forms the favourable half plane for IH,
whereas LHP is the favourable half plane for NH.
Even if we use the anti-neutrino oscillation probabilities,
the same considerations will hold. Therefore, the same 
relation between hierarchy and half-plane holds for both
neutrino and anti-neutrino data. 

\begin{figure}[H]
\vspace{-1.7cm}
\begin{center}
       \epsfig{file=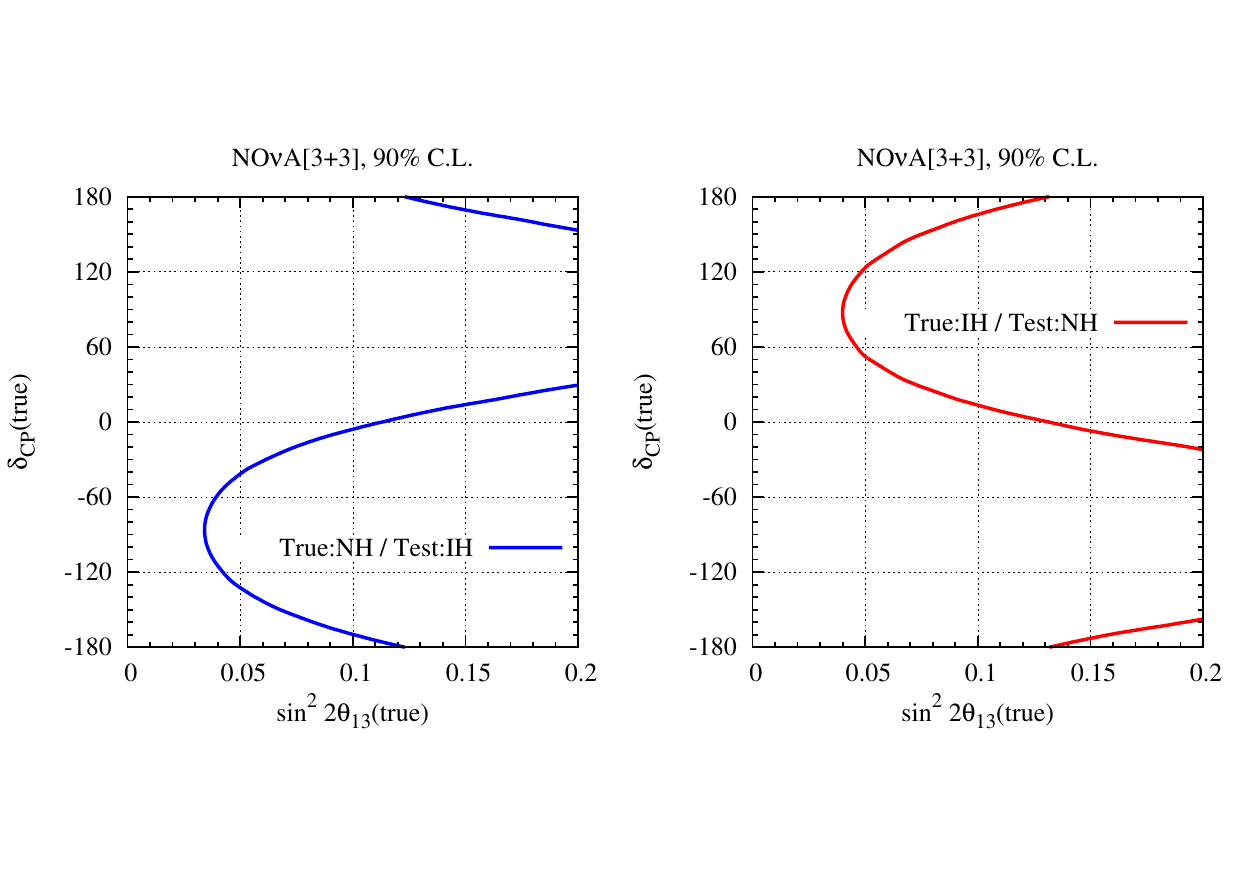,width=1\textwidth}
\end{center}
\vspace{-2.7cm}
\caption{\label{hiernova}\footnotesize (colour online) 
Hierarchy exclusion plots for NO$\nu$A for 3$\nu$+3$\bar\nu$ running
when NH is true (left panel) and when IH is true (right panel)}
\end{figure}

We plot the hierarchy discrimination
ability of \nova in Fig.~\ref{hiernova}. We see that, for
$\sin^2 2 \ty = 0.1$, the
hierarchy can be determined at 90 \% C.L. for the following two combinations:
(NH, $-170^\circ \leq \dcp \leq -10^\circ$) or (IH, $10^\circ \leq \dcp
\leq 170^\circ$). The statistics for the experiment 
are not quite enough to determine the hierarchy for the 
whole favourable half plane for this value of $\ty$.
If $\sin^2 2 \ty = 0.12$, then the hierarchy
can be determined for the whole favoured half plane. It was
shown in Ref.~\cite{hubercpv} that \nova can determine the 
hierarchy for 45 \% of the $\dcp$ range for $\sin^2 2 \ty = 0.1$.

\begin{figure}[H]
\vspace{-1.7cm}
\begin{center}
       \epsfig{file=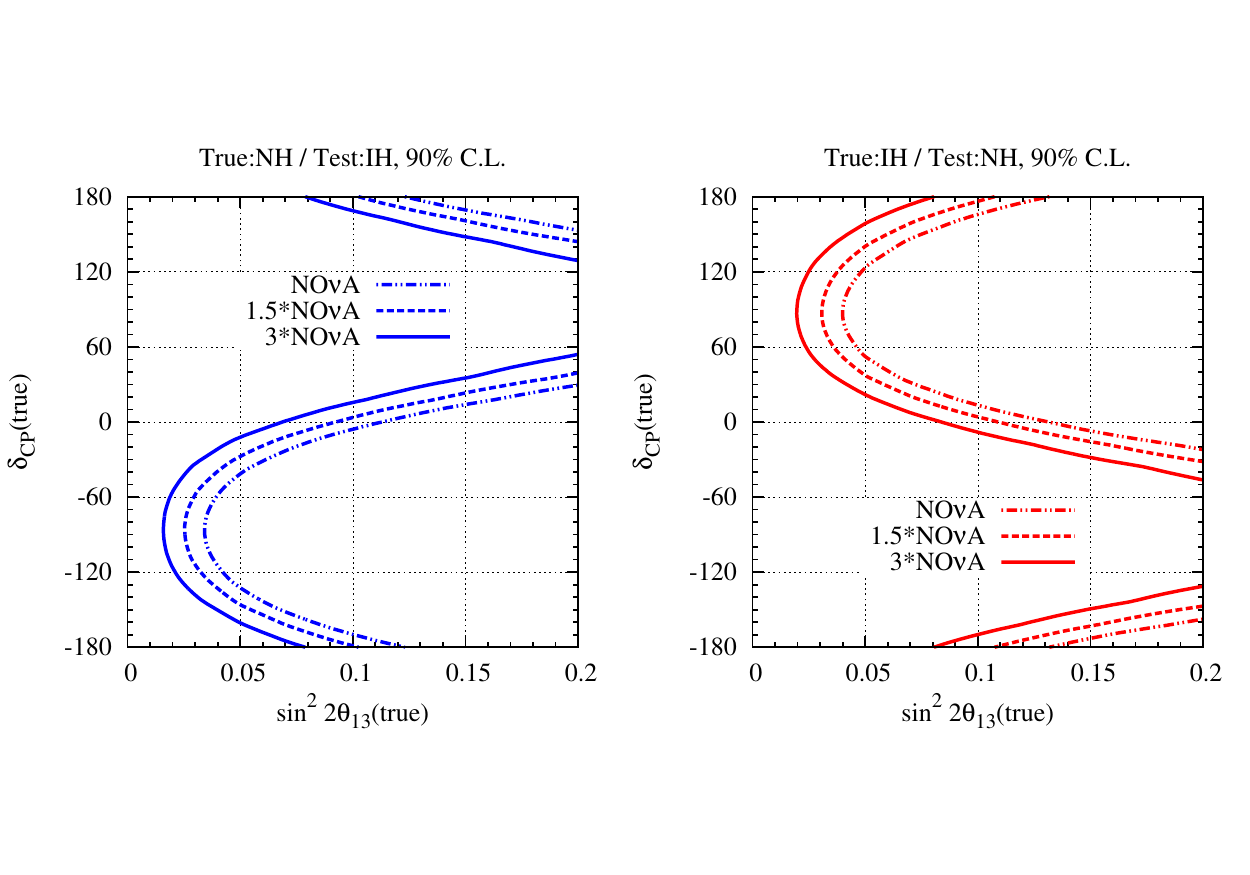,width=1\textwidth}
\end{center}
\vspace{-2.7cm}
\caption{\label{boostnova}\footnotesize (colour online) 
Hierarchy exclusion plots for NO$\nu$A with 
boosted statistics for 3$\nu$+3$\bar\nu$ running when NH is true 
(left panel) and when IH is true (right panel).}
\end{figure}

For smaller values of $\sin^ 2 2 \ty$, one needs larger 
statistics to determine the hierarchy for the whole favourable
half plane. This is illustrated in Fig.~\ref{boostnova}.
With 1.5 times the presently projected statistics of \nova,
one can determine the hierarchy for the whole of the respective favourable
half planes, for both NH and IH, for $\sin^2 2 \theta_{13} = 0.1$. 
Similar conclusions were obtained earlier in Ref.~\cite{nova_tdr}.
If $\dcp$ happens to be in the unfavourable half plane,
even tripling of statistics leads to hierarchy determination only 
for a very small range of $\dcp$. 

\subsection{Resolving the hierarchy-$\dcp$ degeneracy with T2K}

As we demonstrated in the previous subsection, \nova alone can't 
determine the hierarchy if nature chooses one of the unfavourable
combinations (NH, UHP) or (IH, LHP). In this subsection, we explore how data 
from T2K can help in resolving this problem. Since the baseline
of T2K is smaller, the probability peaks at a lower energy and 
hence the flux is designed to peak at a lower energy. Therefore
the matter term $A$ is much smaller for T2K.

 \begin{figure}[H]
\vspace{-1.7cm}
\begin{center}
	\epsfig{file=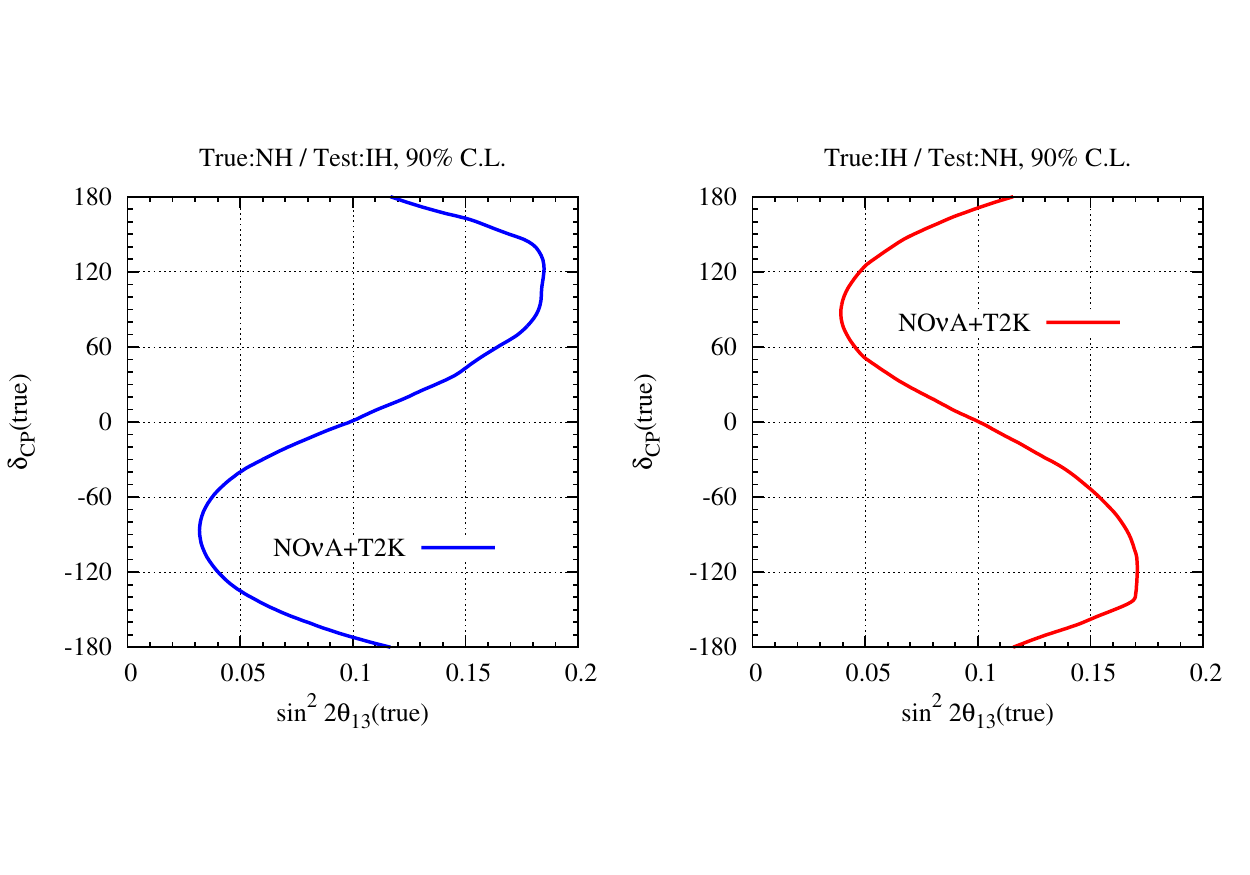,width=1\textwidth}
\end{center}
\vspace{-2.7cm}
\caption{\label{t2knominal}\footnotesize (colour online) 
Hierarchy exclusion plots for NO$\nu$A + T2K with 
nominal statistics when NH is true (left panel) and when IH is true 
(right panel).}
\end{figure}

In Fig.~\ref{t2knominal}, we plot the combined hierarchy exclusion
capability of \nova and T2K.
From this figure we see that, for $\sin^2 2 \ty \leq 0.1$, 
hierarchy determination is not possible for any   
$\dcp$ in the unfavourable half-plane, 
Hence, in our example, we assume that the 
statistics of \nova are 50 \% more than the nominal value and those of
T2K are twice the nominal value.

\begin{figure}[H]
\vspace{-1.7cm}
\begin{center}
	\epsfig{file=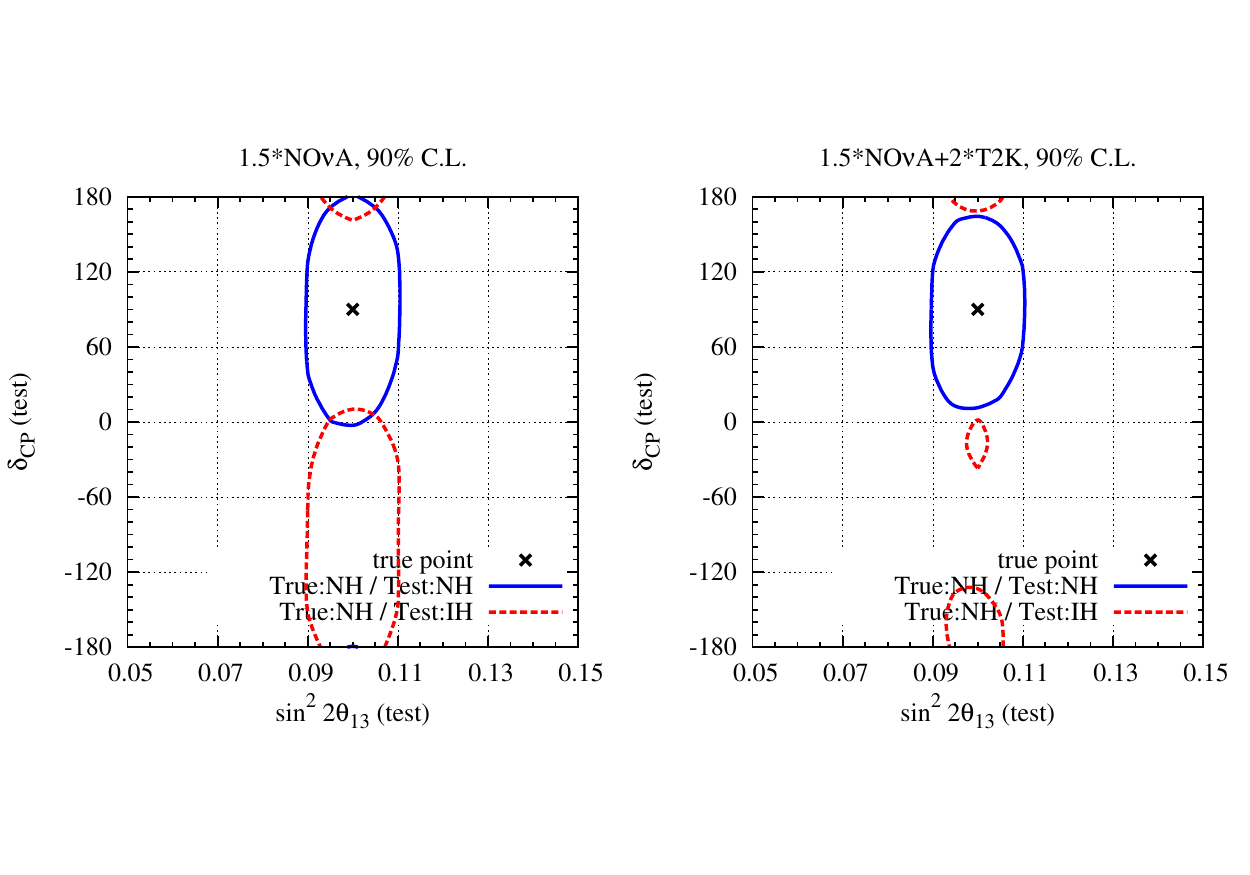,width=1\textwidth}
\end{center}
\vspace{-2.7cm}
\caption{\label{novaNHpiby2}\footnotesize (colour online) 
Allowed $\sin^2 2 \ty$-$\dcp$ plots for 1.5*\nova
(left panel) and 1.5*\nova + 2*T2K (right panel) with true 
$\sin^2 2 \ty = 0.1$ and true $\dcp = 90^\circ$.}
\end{figure}

We illustrate the effect of T2K data on hierarchy determination 
by a set of examples. First we assume that NH is the true hierarchy and the true value
of $\dcp = 90^\circ$, i.e. 
in the unfavourable half plane. In such a situation, \nova
data gives two degenerate solutions in the form of
(NH, $\dcp \approx 90^\circ$) and (IH, $\dcp$ in LHP),
as shown in Fig.~\ref{novaNHpiby2} (left panel). 

\begin{figure}[H]
\vspace{-1.7cm}
\begin{center}
	\epsfig{file=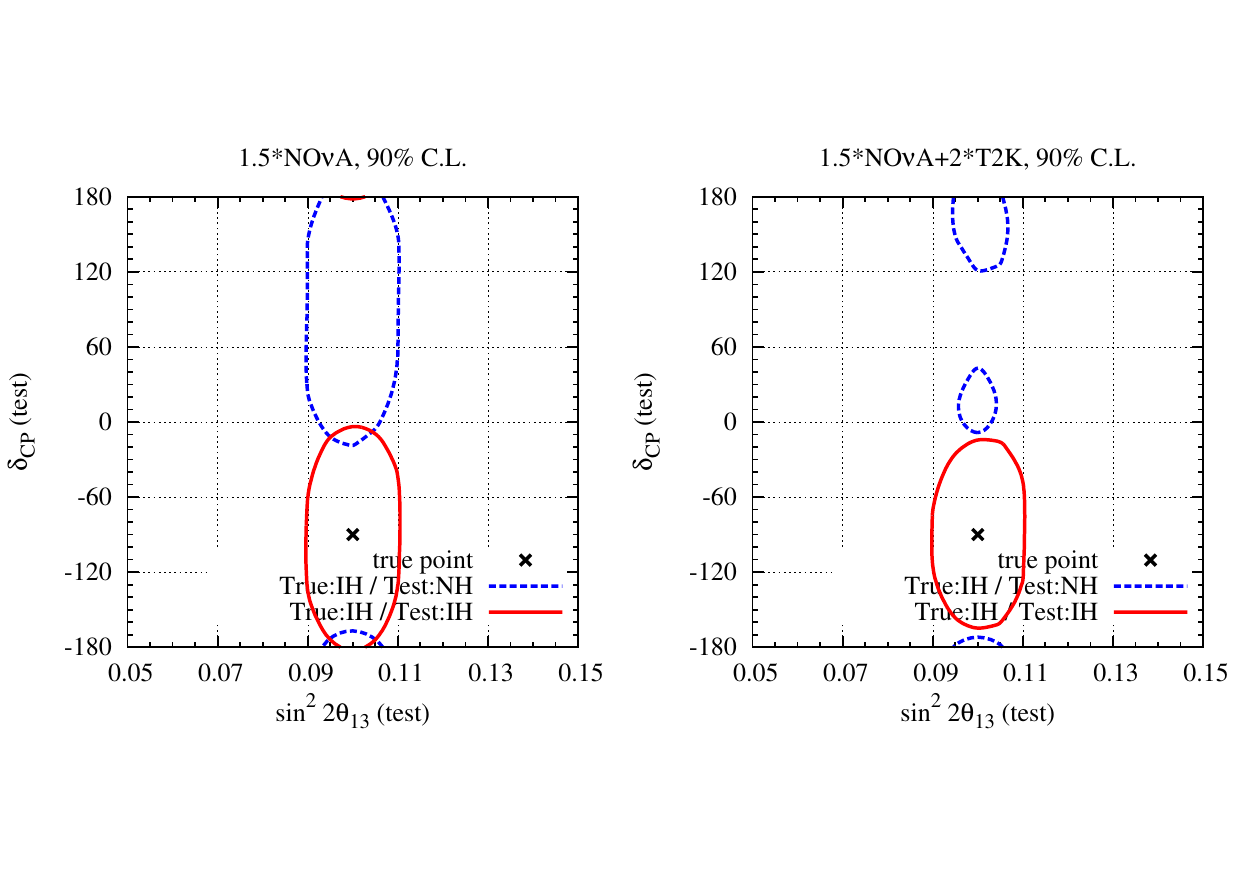,width=1\textwidth}
\end{center}
\vspace{-2.7cm}
\caption{\label{novaIHmpiby2}\footnotesize (colour online) 
Allowed $\sin^2 2 \ty$-$\dcp$ plots for 1.5*\nova
(left panel) and 1.5*\nova + 2*T2K (right panel) with true 
$\sin^2 2 \ty = 0.1$ and true $\dcp = -90^\circ$.}
\end{figure}

But, the addition of T2K data almost rules out the 
(IH,LHP) solution, seen in
the right panel of Fig.~\ref{novaNHpiby2}. It is true
that a very small part of the allowed region is left behind. 
But, comparing the two panels of Fig.~\ref{novaNHpiby2}, 
we see that the addition of T2K data reduces the allowed NH region
only by a small amount whereas the allowed IH region
is drastically reduced. This gives a
very strong indication of which hierarchy is correct. 
Thus the data
of \nova in conjunction with that of T2K can effectively
discriminate against the wrong hierarchy. This holds 
true for the case of IH being the true hierarchy with
$\dcp$ in LHP, illustrated in Fig.~\ref{novaIHmpiby2}.
Figs.~\ref{novaNHpiby2} and~\ref{novaIHmpiby2} are similar
to figures 2 and 3 of Ref. \cite{hubercpv}, 
which are done for the same $\dcp$ values. 
Those figures also show the large shrinkage 
of the wrong hierarchy solution, with the addition of T2K data.
In the following, we will demonstrate that this feature occurs
for all values of $\dcp$(true) in the unfavourable half-plane.

A theoretical analysis of the hierarchy-$\dcp$ degeneracy
resolution, with data from \nova and T2K, was done in 
Ref. \cite{menaparke}. To keep the arguments simple, first
it was assumed that $\theta_{23}$ is maximal and that
$\sin^2 2 \theta_{13}$ is measured accurately by the 
reactor experiments. In such a situation, given a probability
measurement, there exist two degenerate solutions: (correct
hierarchy, correct $\dcp$) and (wrong hierarchy, wrong $\dcp$).
In Ref. \cite{menaparke}, it was shown that, for a given experiment, 
[$\sin({\rm correct}~\dcp) - \sin({\rm wrong}~\dcp)]$ is proportional
to the matter term $A$ for that experiment. For T2K, this
difference is small and is about $0.7$ for $\sin^2 2 \theta_{13}
=0.1$. For \nova it is three times larger. Therefore, the 
wrong $\dcp$ values for T2K data and for \nova data are widely
different. A combined analysis of data from T2K and \nova will
pick out the correct hierarchy and a range of $\dcp$ around
the correct value, provided the statistics from each experiment
are large enough
 
The above idea is illustrated below in Figs.~\ref{chisqnova}
and~\ref{chisqt2k}.
In Fig.~\ref{chisqnova}, we have plotted $\chi^2$ vs $\dcp$(test)
for various true values of $\dcp$ for \nova experiment.
In the left panel, the true 
values of $\dcp$ are all in LHP which is the favourable half-plane
for NH. We find that, except for the CP conserving case of
$\dcp = - 180^\circ$, all the $\chi^2$ are above 9. Hence the wrong
hierarchy can be excluded for most of the values of $\dcp$ in
the favourable half-plane. In the right panel, the true values 
of $\dcp$ are all in UHP, which is the unfavourable half-plane
for NH. And we find that in all cases, the $\chi^2$ becomes
nearly zero (except for $\dcp=0$) for
$-120^\circ \leq \dcp~({\rm test}) \leq -60^\circ$. 
These are the degenerate (wrong hierarchy,
wrong $\dcp$) solutions mentioned above. Hence it is
impossible to rule out the wrong hierarchy if true $\dcp$ is 
in the unfavourable half-plane. 

\begin{figure}[H]
\vspace{-1.7cm}
\begin{center}
	\epsfig{file=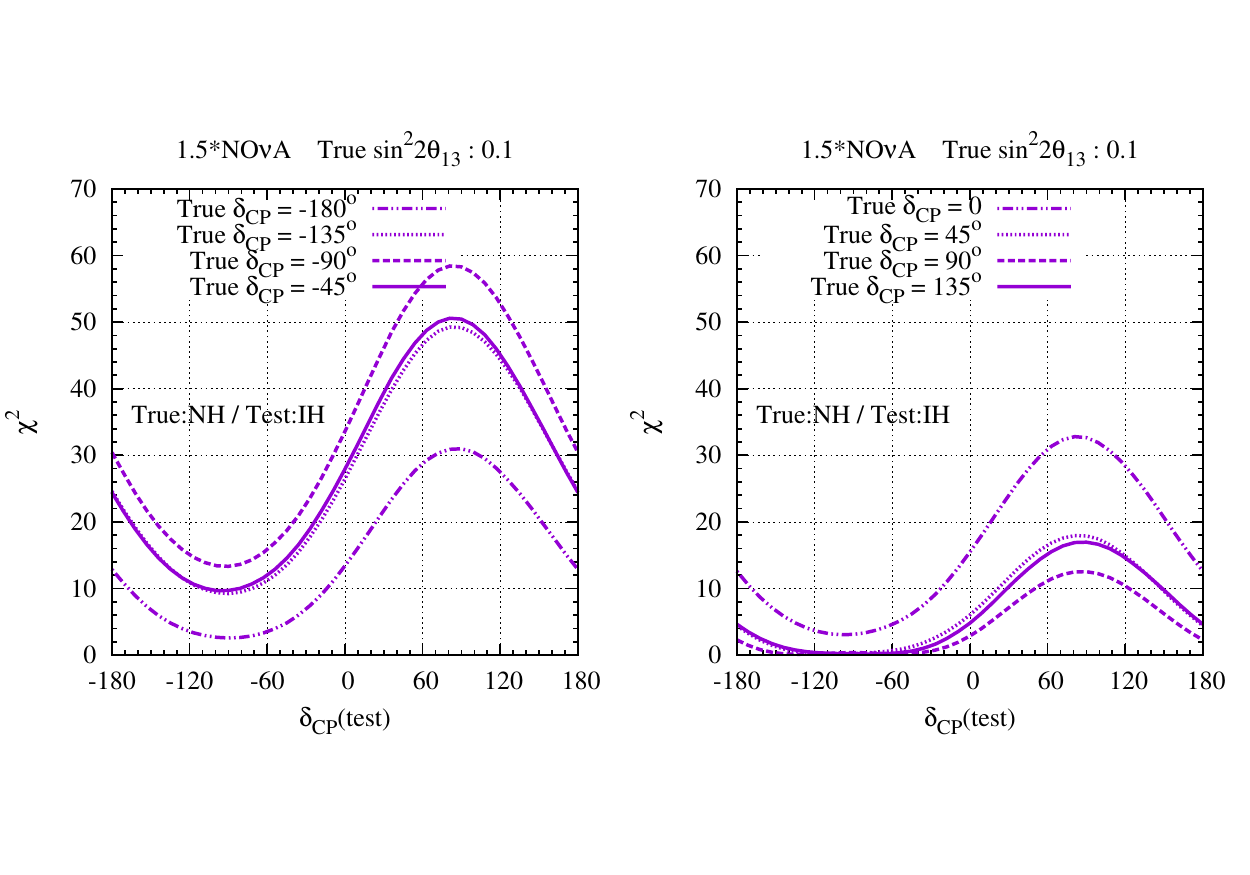,width=1\textwidth}
\end{center}
\vspace{-2.7cm}
\caption{\label{chisqnova}\footnotesize (colour online) 
$\chi^2$ vs. test $\dcp$ for 1.5*\nova. Here true and test $\sin^22\ty=0.1$. 
NH is true and IH is test. Different curves correspond to various
true $\dcp$ in lower half plane (left panel) and upper half plane 
(right panel).}
\end{figure}

In Fig.~\ref{chisqt2k}, we have plotted $\chi^2$ vs $\dcp$(test)
for various true values of $\dcp$ for T2K experiment. 
Once again, the left panel contains 
plots for $\dcp$ in LHP and the right panel the plots for 
$\dcp$ in UHP. From the left panel, we see that T2K can't rule
the wrong hierarchy. This is to be contrasted with \nova case, 
where the wrong hierarchy is ruled out by \nova alone, if $\dcp$ 
is in the favourable half-plane. But, as we see below, T2K data
is crucial for hierarchy discrimination, if $\dcp$ is in the
unfavourable half-plane.

\begin{figure}[H]
\vspace{-1.7cm}
\begin{center}
	\epsfig{file=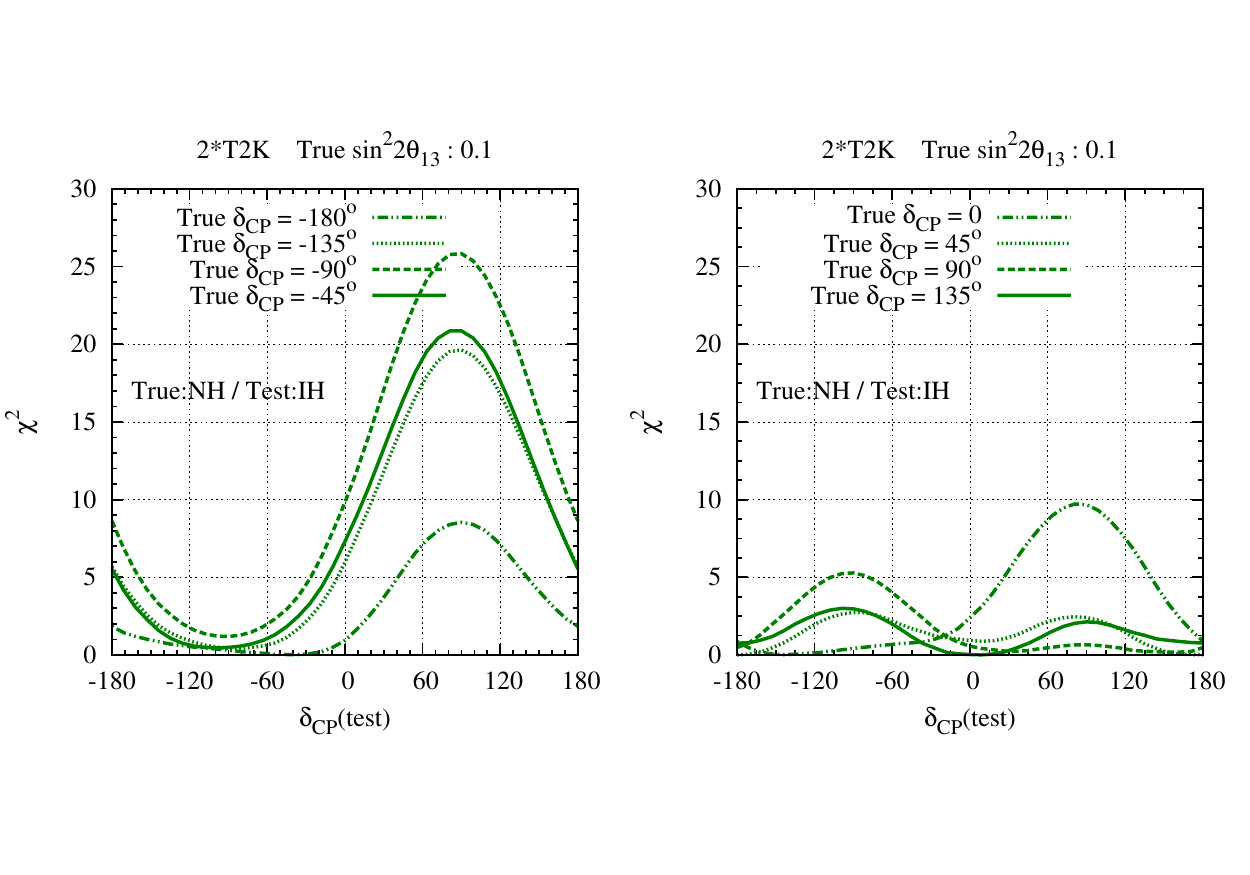,width=1\textwidth}
\end{center}
\vspace{-2.7cm}
\caption{\label{chisqt2k}\footnotesize (colour online) 
$\chi^2$ vs. test $\dcp$ for 2*T2K. Here true and test $\sin^22\ty=0.1$. 
NH is true and IH is test. Different curves correspond to various
true $\dcp$ in lower half plane (left panel) and upper half plane 
(right panel).}
\end{figure}

From the right panel, we see that the degenerate (wrong 
hierarchy, wrong $\dcp$) solution for T2K occurs for 
$\dcp$(test) around $0$ or $\pm 180^\circ$. And in the range $-120^\circ
\leq \dcp~({\rm test}) \leq -60^\circ$, where the degenerate
wrong hierarchy solution for \nova occured, 
the $\chi^2$ for T2K is quite large. 
Because of this wide difference between the $\dcp$ values of the degenerate
(wrong hierarchy, wrong $\dcp$) solutions of \nova data 
and T2K data, together they rule out the wrong hierarchy.

\begin{figure}[H]
\vspace{-1.7cm}
\begin{center}
        \epsfig{file=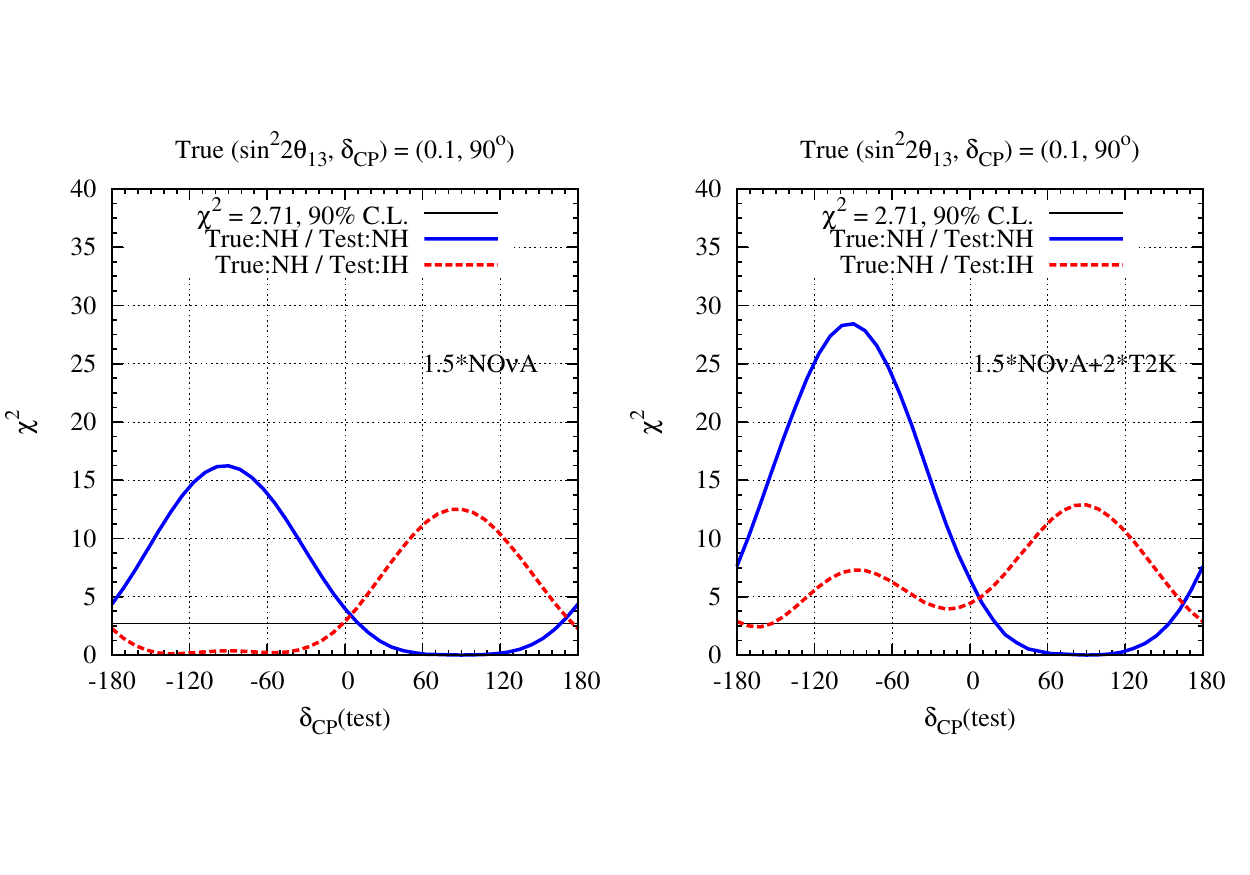,width=1\textwidth}
\end{center}
\vspace{-2.7cm}
\caption{\label{novachiNHpiby2}\footnotesize (colour online)
$\chi^2$ vs $\dcp$(test) plots for 1.5*\nova
(left panel) and 1.5*\nova + 2*T2K (right panel) with true
$\sin^2 2 \ty = 0.1$ and true $\dcp = 90^\circ$.}

\end{figure}
\begin{figure}[H]
\vspace{-1.7cm}
\begin{center}
        \epsfig{file=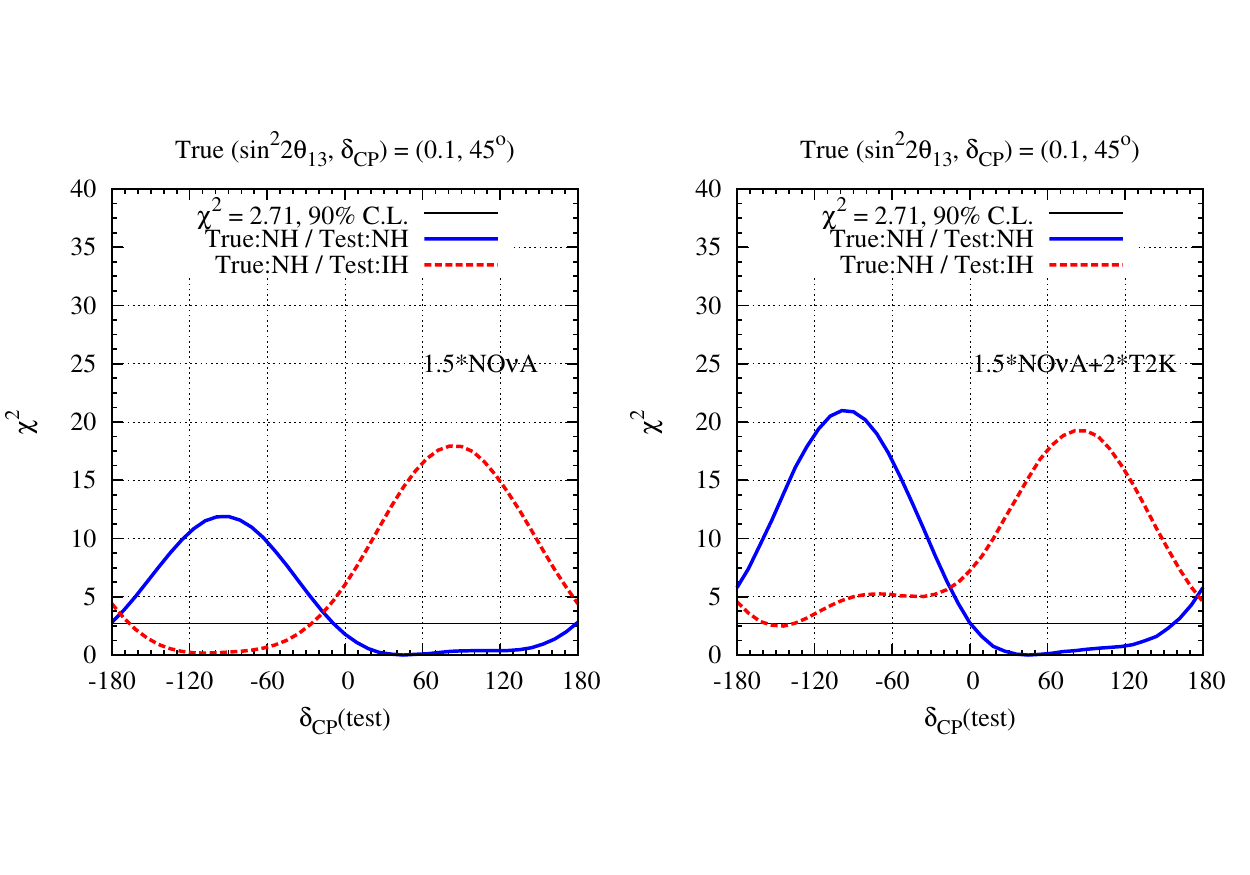,width=1\textwidth}
\end{center}
\vspace{-2.7cm}
\caption{\label{novachiNHpiby4}\footnotesize (colour online)
$\chi^2$ vs $\dcp$(test) plots for 1.5*\nova
(left panel) and 1.5*\nova + 2*T2K (right panel) with true
$\sin^2 2 \ty = 0.1$ and true $\dcp = 45^\circ$.}
\end{figure}
\begin{figure}[H]
\vspace{-1.7cm}
\begin{center}
        \epsfig{file=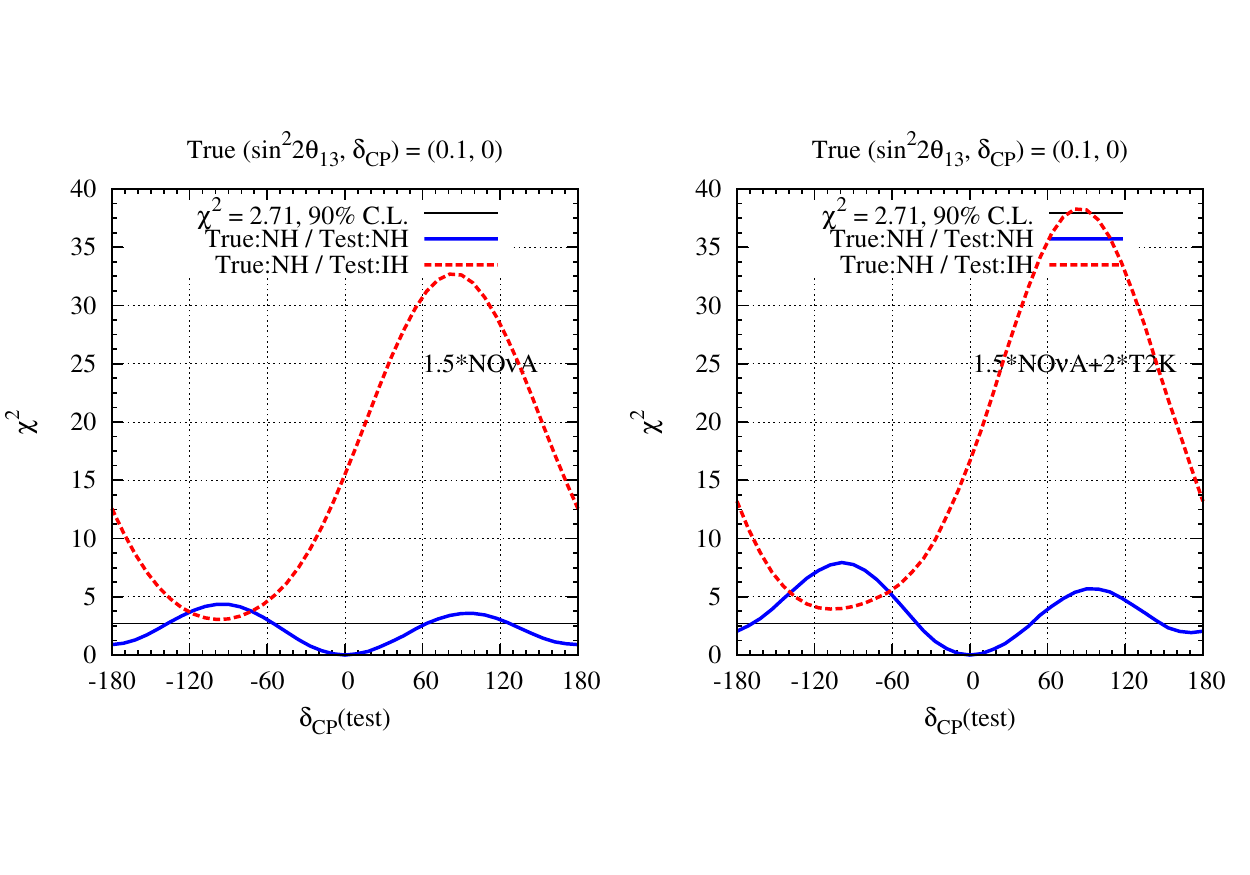,width=1\textwidth}
\end{center}
\vspace{-2.7cm}
\caption{\label{novachiNH0}\footnotesize (colour online)
$\chi^2$ vs $\dcp$(test) plots for 1.5*\nova
(left panel) and 1.5*\nova + 2*T2K (right panel) with true
$\sin^2 2 \ty = 0.08$ and true $\dcp = 0$
(systematics included).}
\end{figure}
\begin{figure}[H]
\vspace{-1.7cm}
\begin{center}
        \epsfig{file=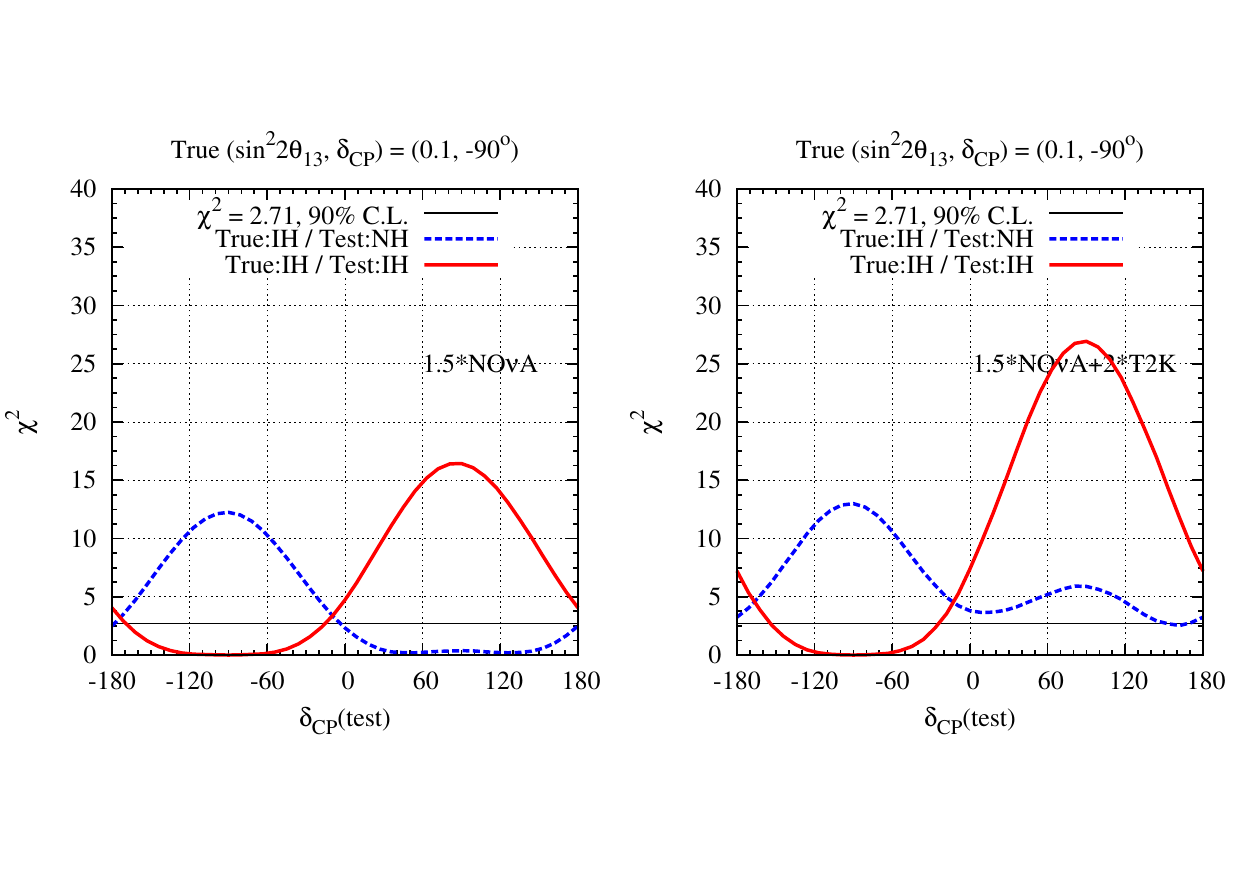,width=1\textwidth}
\end{center}
\vspace{-2.7cm}
\caption{\label{novachiIHmpiby2}\footnotesize (colour online)
$\chi^2$ vs $\dcp$(test) plots for 1.5*\nova
(left panel) and 1.5*\nova + 2*T2K (right panel) with true
$\sin^2 2 \ty = 0.1$ and true $\dcp = -90^\circ$.}
\end{figure}
\begin{figure}[H]
\vspace{-1.7cm}
\begin{center}
        \epsfig{file=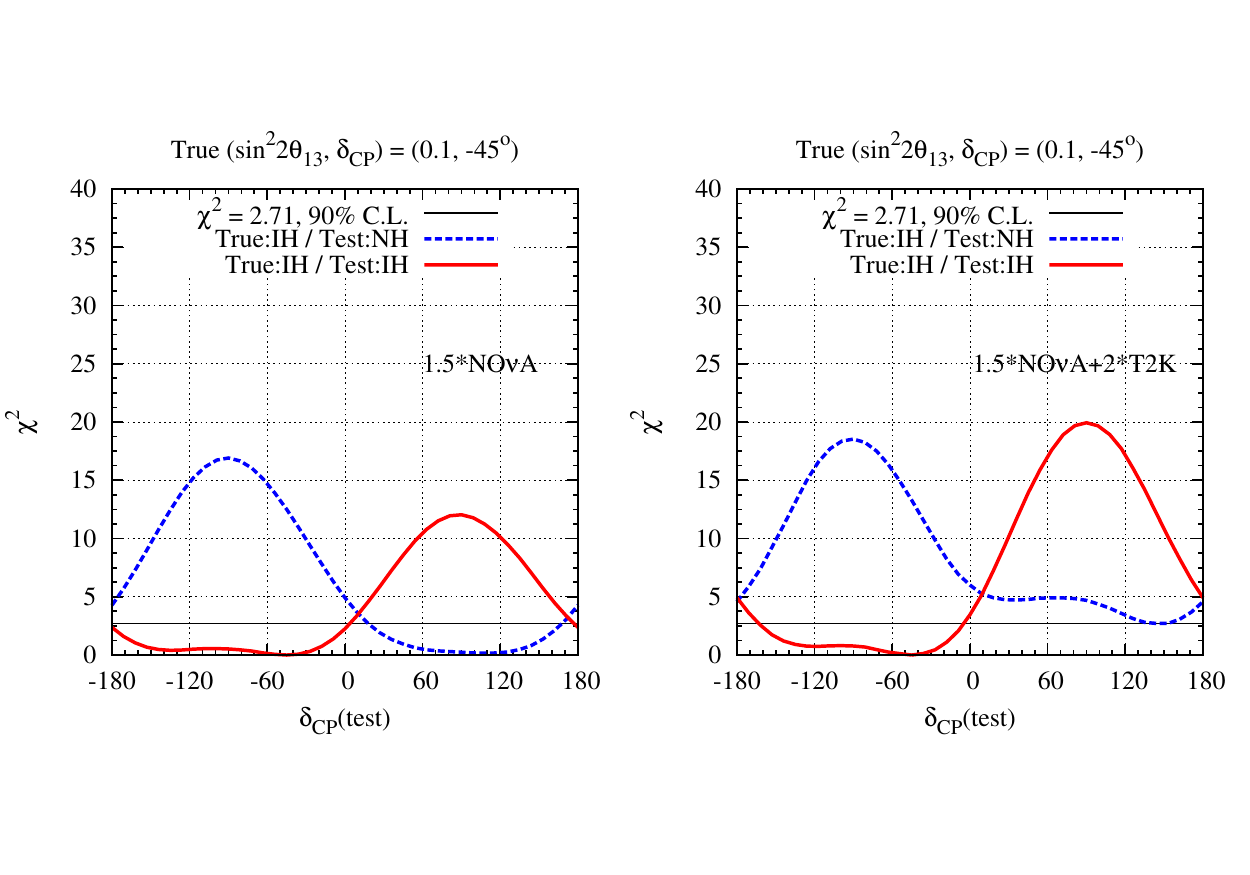,width=1\textwidth}
\end{center}
\vspace{-2.7cm}
\caption{\label{novachiIHmpiby4}\footnotesize (colour online)
$\chi^2$ vs $\dcp$(test) plots for 1.5*\nova
(left panel) and 1.5*\nova + 2*T2K (right panel) with true
$\sin^2 2 \ty = 0.1$ and true $\dcp = -45^\circ$.}
\end{figure}
\begin{figure}[H]
\vspace{-1.7cm}
\begin{center}
        \epsfig{file=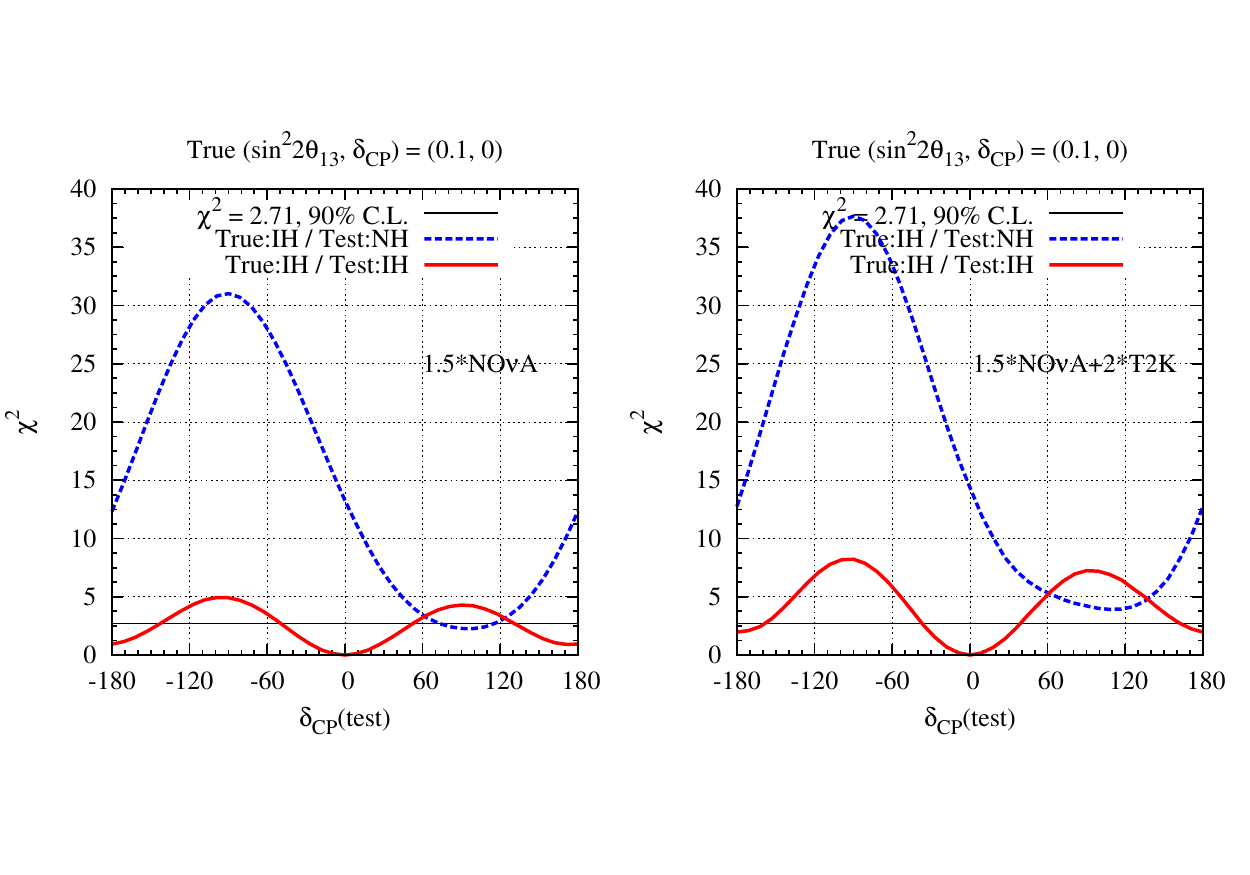,width=1\textwidth}
\end{center}
\vspace{-2.7cm}
\caption{\label{novachiIH0}\footnotesize (colour online)
$\chi^2$ vs $\dcp$(test) plots for 1.5*\nova
(left panel) and 1.5*\nova + 2*T2K (right panel) with true
$\sin^2 2 \ty = 0.1$ and true $\dcp = 0$.}
\end{figure}

We illustrate this hierarchy discriminating power 
for a few cases where true value of $\dcp$
is in the unfavourable half plane. 
Figs.~\ref{novachiNHpiby2},~\ref{novachiNHpiby4}
and~\ref{novachiNH0} show the $\chi^2$  
plots for $\dcp = 90^\circ, 45^\circ \ {\rm and} \ 0$
respectively, with NH as the true hierarchy.
The left panel shows $\chi^2$ for 1.5*\nova alone whereas 
the right panel shows the $\chi^2$ for 1.5*\nova+2*T2K. 
These plots show $\chi^2$ for the two cases where
the true and test hierarchies are the same and are opposite.
In these plots, we have marginalised over $\sin^2 2\ty$. 
In the left panel of Fig.~\ref{novachiNHpiby2}, 
there is a large allowed region of $\dcp$(test) in the wrong 
half-plane, if the test hierarchy is the wrong hierarchy. In
the right panel, this region is almost completely ruled out, 
with the addition of T2K data. 
There is a just a small region, centered around
$\dcp$(test) $\approx 180^\circ$, where the $\chi^2$ dips
just below $2.71$, the cut-off for $90 \%$ C.L. 
We see very similar features for
true $\dcp = 45^\circ$ in Fig.~\ref{novachiNHpiby4} and
for true $\dcp = 0$ in Fig.~\ref{novachiNH0}.
Essentially identical features are seen
for the case where IH is true hierarchy in Fig.~\ref{novachiIHmpiby2}
with true $\dcp=-90^\circ$, Fig.~\ref{novachiIHmpiby4} with true $\dcp
= - 45^\circ$ and Fig.~\ref{novachiIH0} with true $\dcp=0$.

\begin{figure}[H]
\vspace{-1.7cm}
\begin{center}
	\epsfig{file=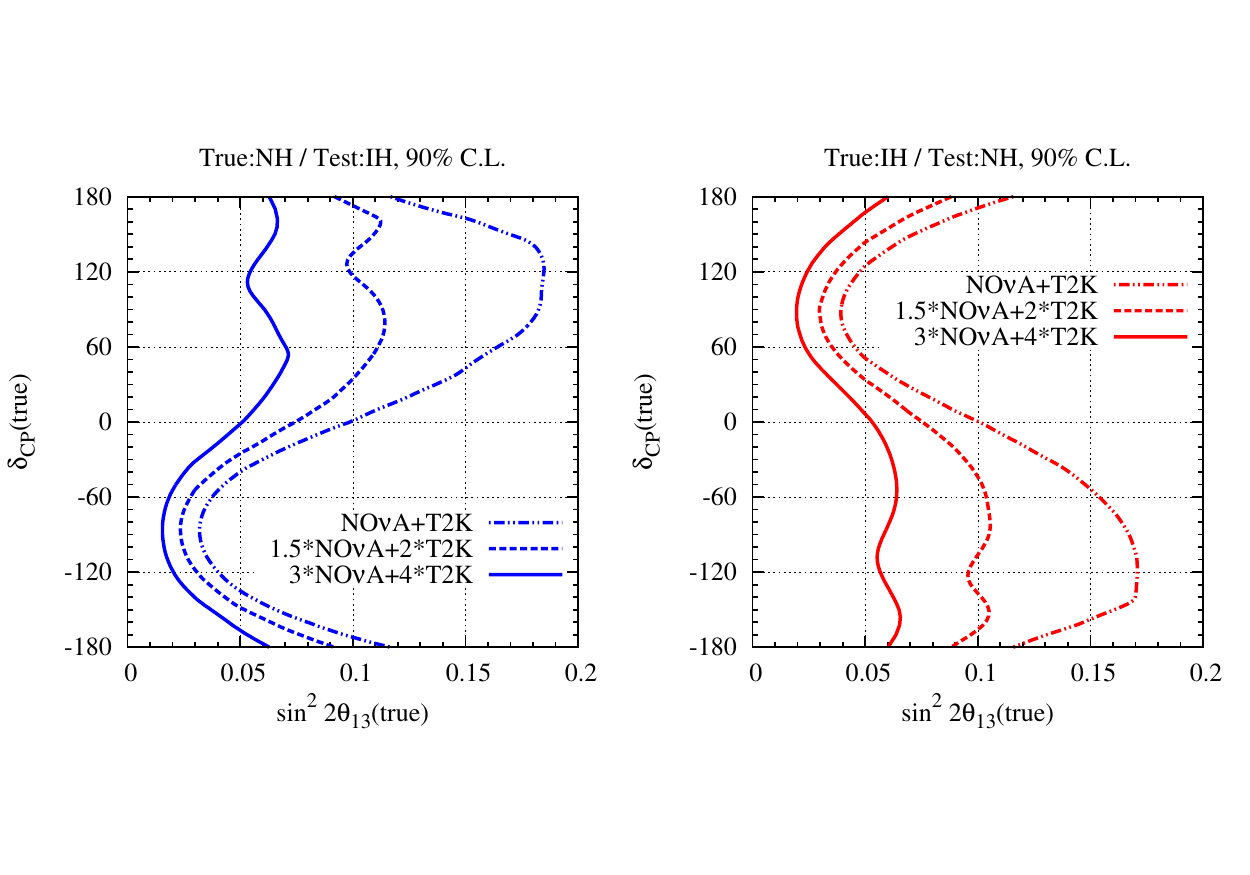,width=1\textwidth}
\end{center}
\vspace{-2.7cm}
\caption{\label{hiernovat2k}\footnotesize (colour online) Hierarchy exclusion plots for 
combined data from NO$\nu$A and T2K with various boosts in statistics
when NH is true (left panel) and when IH is true (right panel).}
\end{figure}

Finally we consider how hierarchy sensitivity improves
with increasing statistics. We consider
three scenarios:
\begin{itemize}
\item
T2K will have a 5 year neutrino run with its design luminosity
and \nova will run according to its present plan.
\item 
T2K will have twice the above statistics and \nova will have
1.5 times its designed statistics. 
\item 
T2K will have four times the above statistics and \nova will have
thrice its designed statistics. 
\end{itemize}
The exclusion plots are given in Fig.~\ref{hiernovat2k}. For
all points to the right of the contours, the wrong hierarchy
can be ruled out. 

In the left panel we assumed NH is the true hierarchy and in
the right panel we assumed IH is the true hierarchy. We see 
that increasing the statistics from nominal values to 
1.5*\nova + 2*T2K dramatically improves the 
ability to rule out the wrong hierarchy, if $\dcp$(true)
is in the unfavourable half-plane. Further improvement 
occurs if the statistics are increased even more.
In particular, if $\sin^2 2 \ty = 0.1$ 
\cite{globalfit_t13,fogli2012}
the hierarchy can be essentially established at $90 \%$
C.L., for any true value of $\dcp$, with 1.5 times the 
designed statistics of \nova and twice the designed 
statistics of T2K. This point was noted previously in Ref.~\cite{nova_tdr}.



It is evident now that an experiment that can exclude the 
wrong $\dcp$ plane effectively can be of great help in determining 
hierarchy when run in conjunction with NO$\nu$A. 
We saw that T2K, with a short baseline and smaller matter effects,
has such properties. We now inquire whether 
having an experiment with a baseline shorter than T2K,
such as C2F, which is 130 km long, can help. 
For such a short baseline, $\pmue$ is maximum
at $E= 0.25$ GeV. At such energies, the matter term $A$ is 
very small. 

\begin{figure}[H]
\vspace{-1.7cm}
\begin{center}
	\epsfig{file=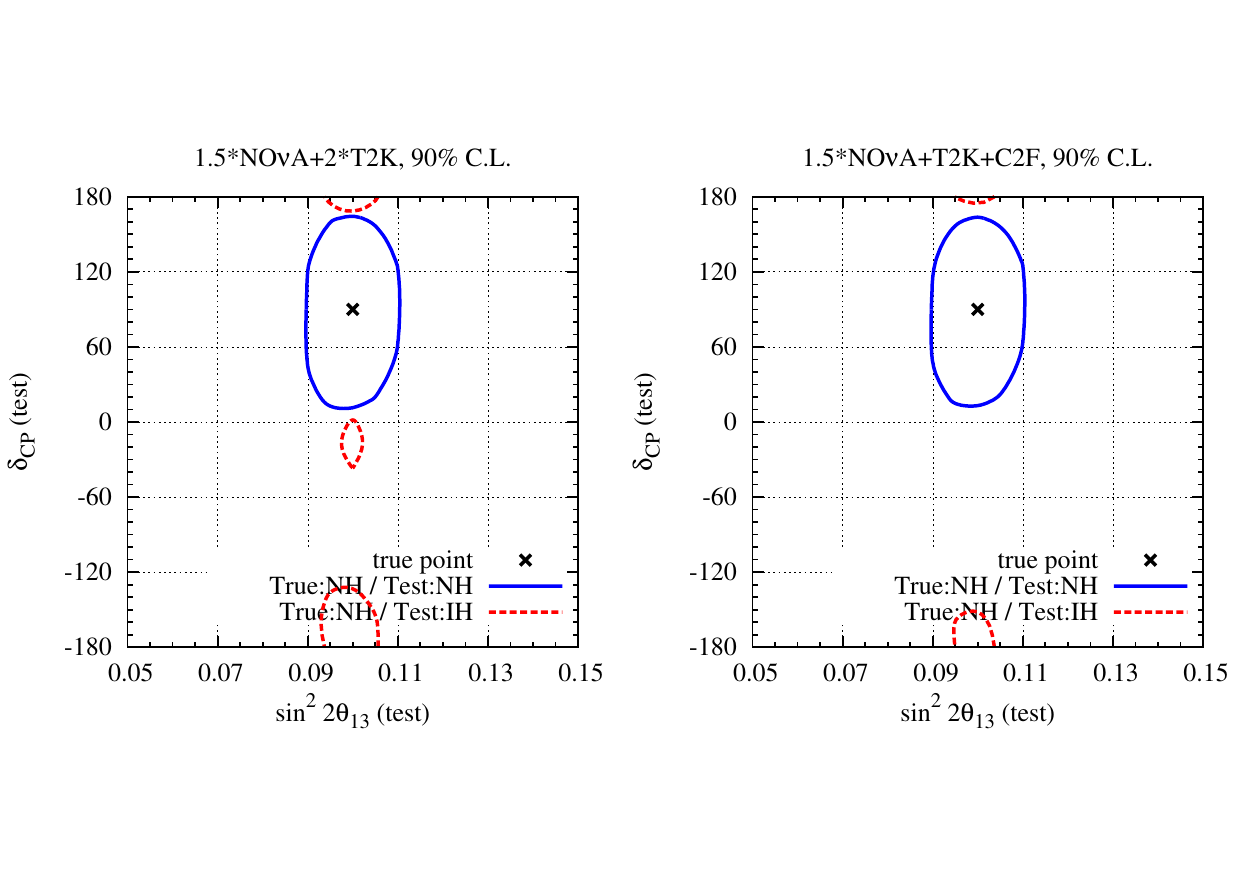,width=1\textwidth}
\end{center}
\vspace{-2.7cm}
\caption{\label{peanutsNHC2F90}\footnotesize (colour online)  
Allowed $\sin^2 2 \ty$-$\dcp$ plots for 1.5*\nova + 2*T2K
(left panel) and 1.5*\nova + T2K + C2F (right panel) with true 
$\sin^2 2 \ty = 0.1$ and true $\dcp = 90^\circ$.}
\end{figure}

To make a just comparison in terms of cost, we assume C2F to have the
same beam power and detector size as that of T2K
and 3 years each of $\nu$ and $\bar{\nu}$ running.
We consider two scenarios.
NO$\nu$A with 1.5 times its designed statistics and T2K with twice its designed statistics 
(scenario A) against NO$\nu$A with 1.5 times its designed statistics 
and T2K and C2F with their nominal designed statistics (scenario B).
In Fig.~{\ref{peanutsNHC2F90}},
we compare the ability of scenario A (left panel) and scenario B 
(right panel) to exclude the wrong hierarchy - 
wrong $\dcp$ region. The two panels are essentially identical. 
We found that scenarios A and B give the same allowed regions 
for all true values of $\dcp$ in
the unfavourable half plane. 
Therefore, a shorter baseline experiment (L $\sim$ 130 km) 
will not help in hierarchy determination.

\section{Measuring $\dcp$ with $\pmue$}

\subsection{$\dcp$ measurement with T2K alone}

In the previous section, we discussed the capability of \nova and T2K to 
determine the mass hierarchy. We now turn our attention to the measurement 
of $\dcp$. Because of the hierarchy-$\dcp$ degeneracy, the determination of 
these two quantities go hand in hand. Matter effects, which are 
hierarchy-dependent, induce 
a CP-like change in the oscillation probabilities. Therefore, it is 
expected that the effects of these two parameters can be disentangled by 
choosing baselines and energies where matter effects are small. Thus, a 
natural choice for accurate measurement of $\dcp$ seems to be an experiment 
with a short baseline and low energy, like T2K or C2F. But, here
we demonstrate that $\dcp$ {\it can not} be measured in such experiments
without first determining the hierarchy. 
For the purpose of this demonstration, in this subsection alone, 
we will assume that T2K will have {\it equal three year runs} in 
neutrino and anti-neutrino modes. This is done because such runs 
have the best capability to determine $\dcp$.
However, even in such a case, $\dcp$ can't be determined without
first determining the hierarchy.

In the following, we present `allowed $\dcp$' graphs. In generating
these, we have kept $\sin^2 2 \ty$ fixed at $0.1$.
The graphs are plotted in the 
true $\dcp$-test $\dcp$ plane. For every true value of $\dcp$, we indicate the 
range in test $\dcp$ that can be excluded at 90\% C.L. The plots have been 
shown for both true and wrong hierarchies. 
The dotted range, defined by $\chi^2 \leq 2.71$, shows the values of test $\dcp$ 
that are compatible with the data, generated with $\dcp$(true) as 
input. For a given true value of $\dcp$, 
the error in measuring $\dcp$ is indicated by the spread of the dotted range 
along that $\dcp$(true) vertical line.

\begin{figure}[H]
\vspace{-1.7cm}
\begin{center}
	\epsfig{file=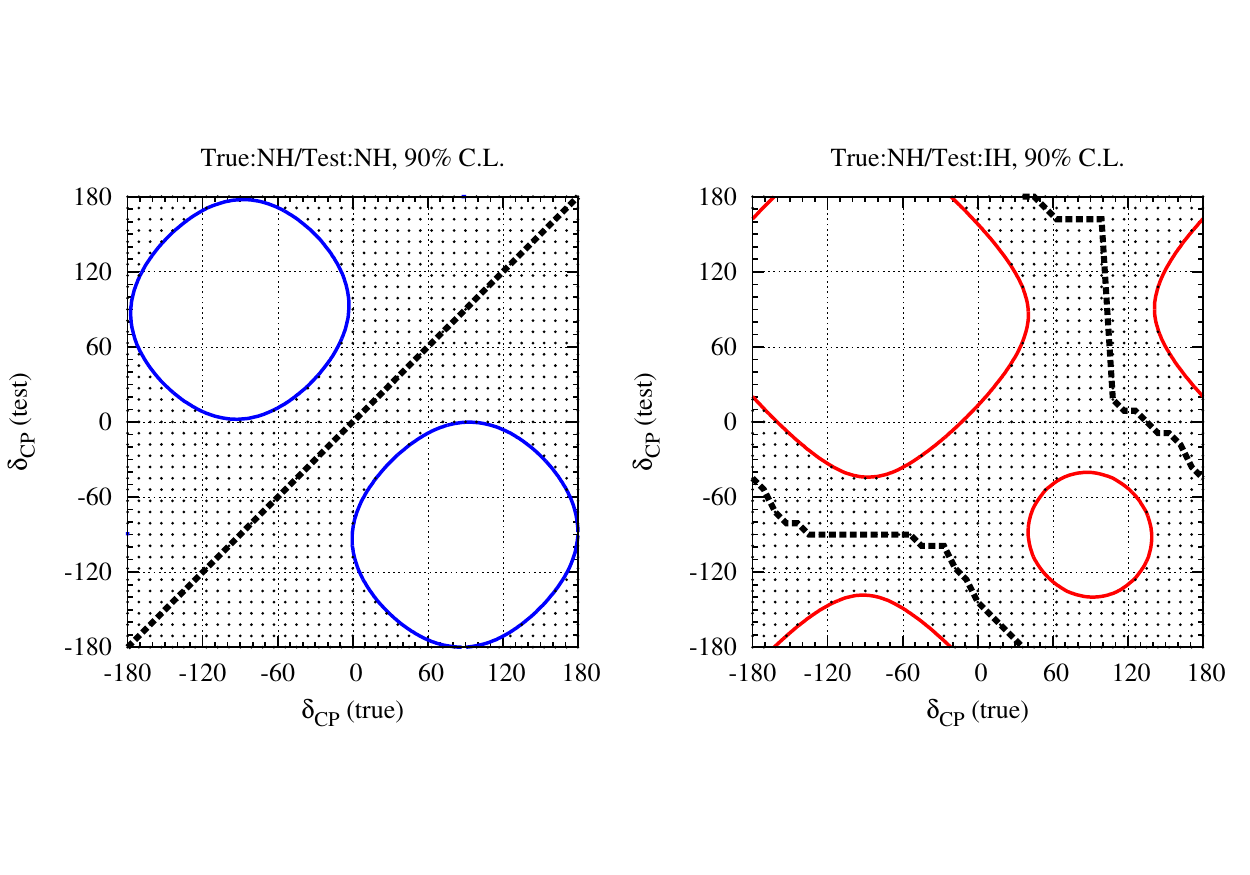,width=1\textwidth}
\end{center}
\vspace{-2.7cm}
\caption{\label{t2kdcp}\footnotesize (colour online) 
Allowed $\dcp$ plots for T2K. Here NH is true. True and test
$\sin^2 2\ty = 0.1$. Test hierarchy is normal (left panel) and 
inverted (right panel).}
\end{figure}

\begin{figure}[H]
\vspace{-1.7cm}
\begin{center}
	\epsfig{file=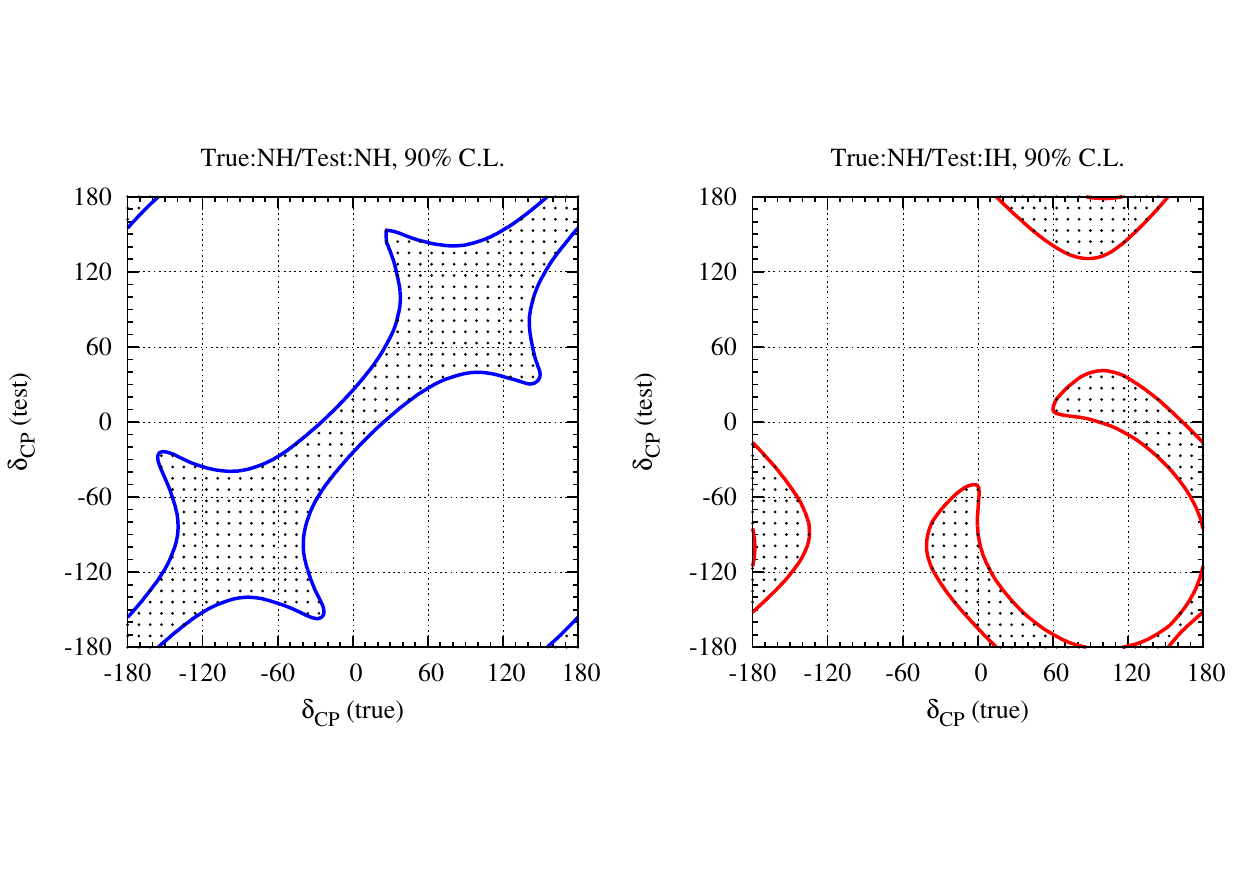,width=1\textwidth}
\end{center}
\vspace{-2.7cm}
\caption{\label{t2k10dcp}\footnotesize (colour online) Allowed $\dcp$ plots 
for 10*T2K. Here NH is true. True and test
$\sin^2 2\ty = 0.1$. Test hierarchy is normal (left panel) and 
inverted (right panel).}
\end{figure}

Figure~(\ref{t2kdcp}) shows the allowed $\dcp$ plot for T2K.
The points on the thick dashed line in this figure correspond
to the values of $\dcp$(test) for which $\chi^2$ is minimum. 
If the test hierarchy is the same as the true hierarchy, then
the $\chi^2$ minimum occurs for $\dcp$(test) = $\dcp$(true) and
the allowed range of test $\dcp$ is around true $\dcp$.
But, if the test hierarchy is the wrong hierarchy, then the
minimum of $\chi^2$ occurs for $\dcp$(test) $\neq$ $\dcp$(true)
and, in general, these two points are widely separated.
This already gives a hint that 
an accurate measurement of $\dcp$ is not possible without first
determining the hierarchy. This point is made more dramatic, when
we consider the situation with more data from T2K. Fig.~\ref{t2k10dcp}
shows the allowed $\dcp$ plot for 10 times the statistics of T2K.
For $\dcp$ (true) in the middle of the favourable half plane
($-140^\circ \leq \dcp \leq -40^\circ$), the wrong hierarchy 
solution is ruled out. Thus both the hierarchy and 
the correct range of $\dcp$ are simultaneously determined. 
For all other values of $\dcp$ (true), we get a wrong value of 
$\dcp$, if we assume the wrong hierarchy.
For example, we see from the right panel of Fig.~\ref{t2k10dcp}, 
for true $\dcp = -30^\circ$, we find that $-130^\circ \leq \dcp({\rm test})
\leq -70^\circ$, when the test hierarchy is the wrong hierarchy.
Similarly for true $\dcp = + 60^\circ$, we find $140^\circ \leq
\dcp({\rm test}) \leq 200^\circ(=-160^\circ)$. In particular, if true $\dcp$
is $-10^\circ$, close to the CP conserving value $0$, we have
$-150^\circ \leq  \dcp({\rm test}) \leq -50^\circ$, encompassing 
maximal CP violation. The situation is similar for true $\dcp=-170^\circ$. 
Conversely, for true $\dcp = 90^\circ$,
we have two allowed regions between $0$ to $40^\circ$ and $140^\circ$ to $180^\circ$, 
both of which
are close to CP conservation. This figure makes it clear that 
it is impossible to have a measurement of $\dcp$ if we do not
know the correct hierarchy. In fact, we are likely to get 
a completely misleading estimate of $\dcp$ if we assume
the wrong hierarchy. The corresponding figures for C2F experiment
show similar features.

\subsection{$\dcp$ measurement with T2K and \nova}

\begin{figure}[H]
\vspace{-1.7cm}
\begin{center}
	\epsfig{file=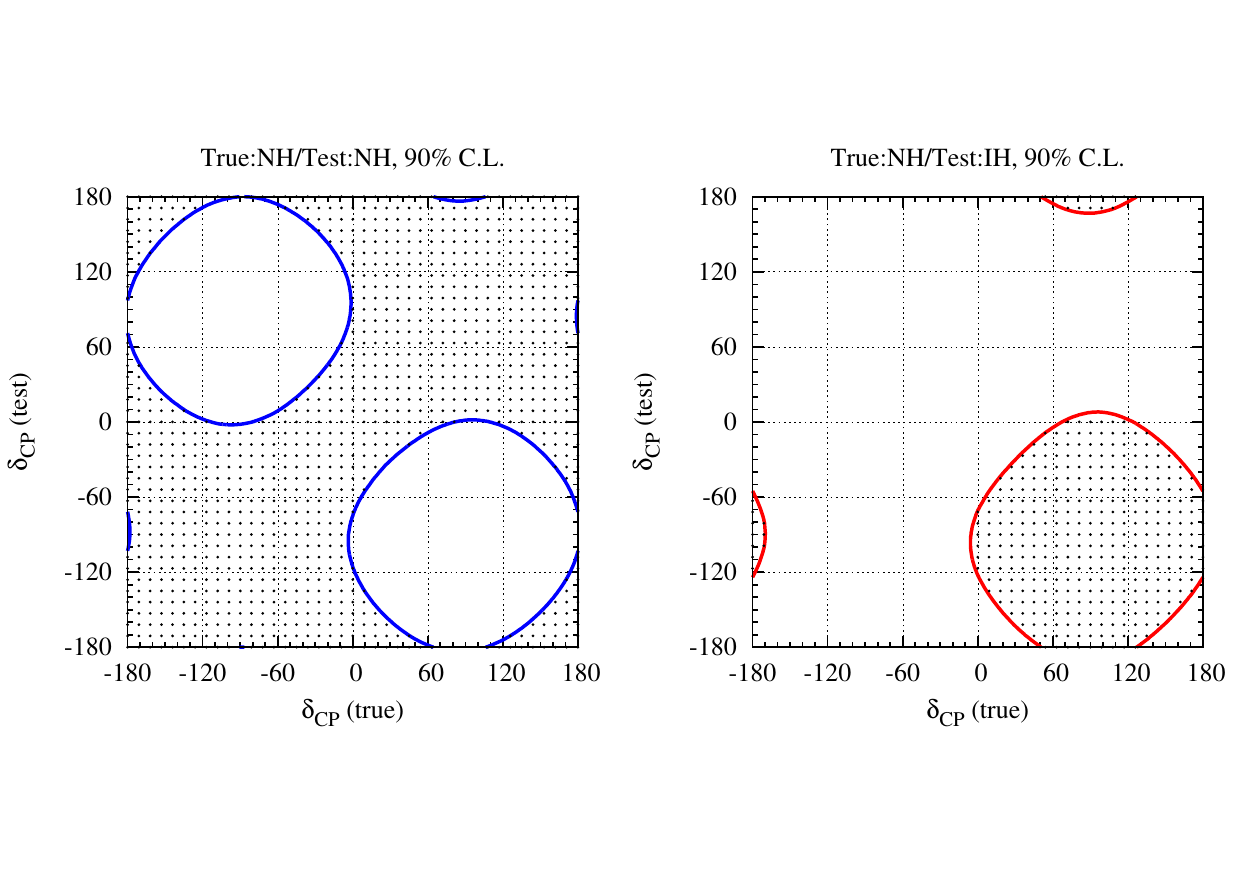,width=1\textwidth}
\end{center}
\vspace{-2.7cm}
\caption{\label{novadcp}\footnotesize (colour online) 
Allowed $\dcp$ plots for \nova. Here NH is true. True and test
$\sin^2 2\ty = 0.1$. Test hierarchy is normal (left panel) and 
inverted (right panel).}
\end{figure}

In this subsection, we consider the $\dcp$ measuring capability of 
\nova and T2K together.
Here we revert back to the original assumption that T2K will run
in neutrino mode only for 5 years.
Fig.~\ref{novadcp}, shows the allowed 
$\dcp$ plot of \nova, assuming NH is true.
If the test hierarchy is the true hierarchy,
the allowed range of $\dcp$ will surround true $\dcp$. If the
test hierarchy is the wrong hierarchy we obtain a large allowed 
range with $\dcp$ far from the true value.

\begin{figure}[H]
\vspace{-1.7cm}
\begin{center}
	\epsfig{file=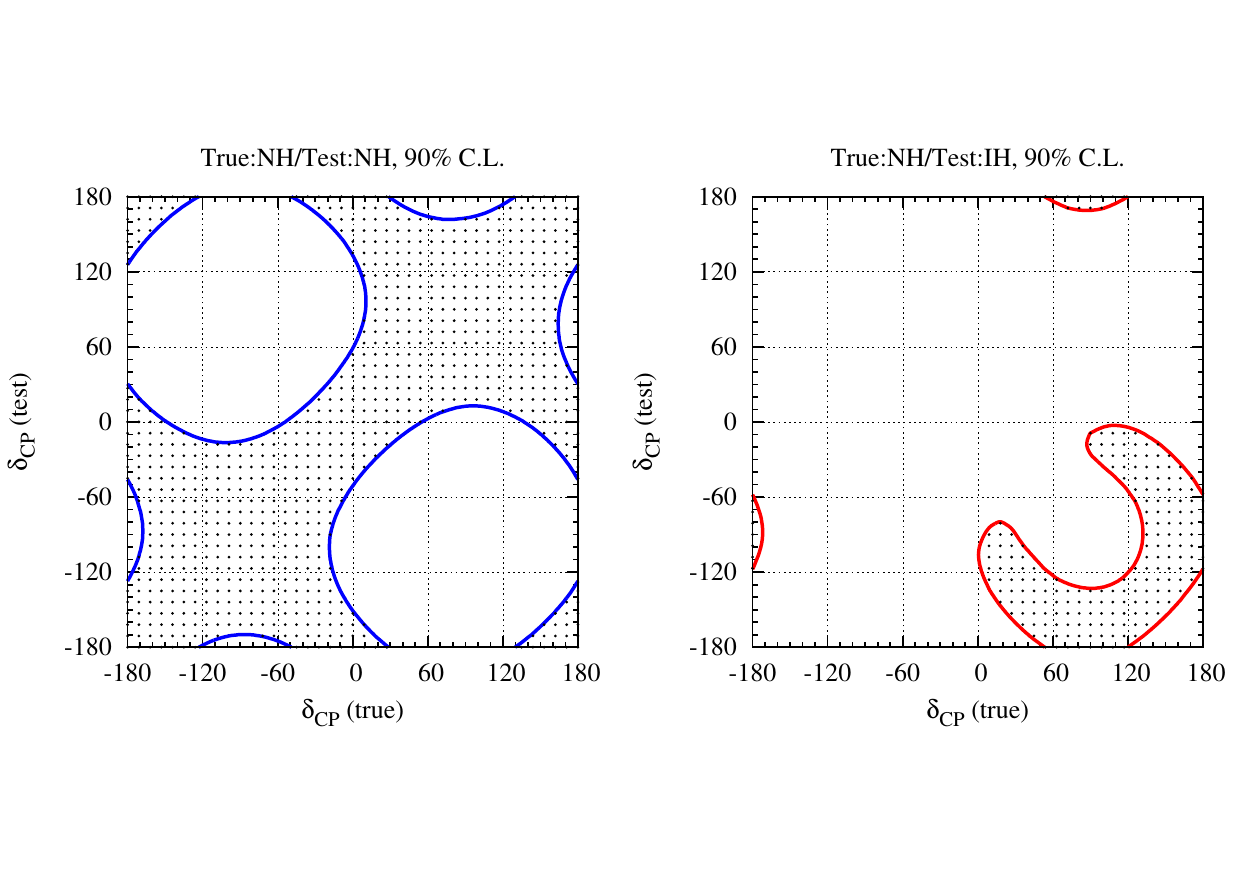,width=1\textwidth}
\end{center}
\vspace{-2.7cm}
\caption{\label{novat2kdcpNH}\footnotesize (colour online) 
Allowed $\dcp$ plots for \nova + T2K. Here NH is true. True and test
$\sin^2 2\ty$ is 0.1. Test hierarchy is normal (left panel) and 
inverted (right panel).}
\end{figure}

Fig.~\ref{novat2kdcpNH} shows the allowed $\dcp$ plot
for \nova and T2K together. In the left panel, the allowed
range $\dcp$ for the true hierarchy is shown. We see that
this range is mostly in the correct half-plane.  
For wrong hierarchy, shown in the right panel, the large allowed 
region in wrong half plane is reduced, but a substantial region
is still allowed. For the case where IH is the true hierarchy,
similar features occur.

\begin{figure}[H]
\vspace{-1.7cm}
\begin{center}
	\epsfig{file=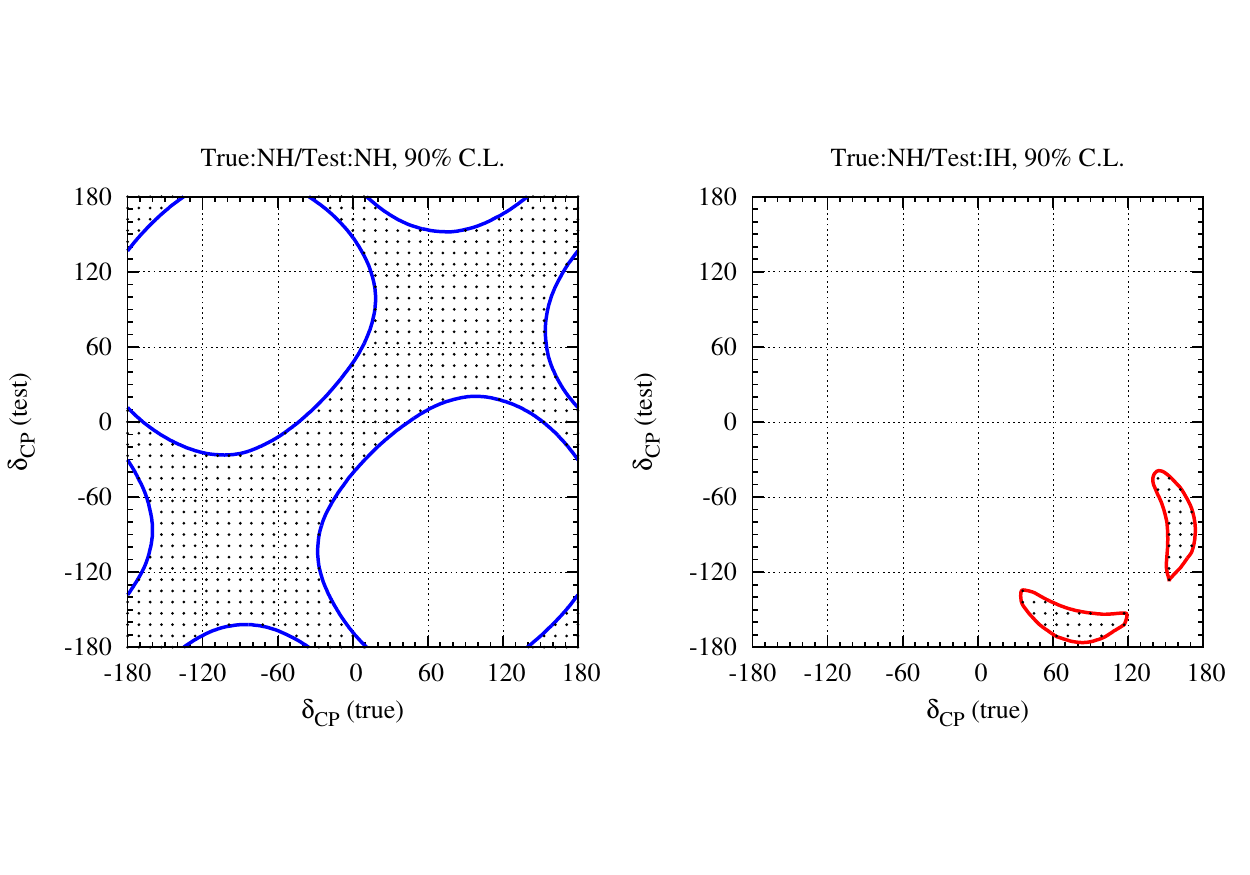,width=1\textwidth}
\end{center}
\vspace{-2.7cm}
\caption{\label{1.5nova2t2kdcp-sysNH}\footnotesize (colour online) 
Allowed $\dcp$ plots for 1.5*\nova + 2*T2K. Here NH is true. True and test
$\sin^2 2\ty = 0.1$. Test hierarchy is normal (left panel) and 
inverted (right panel).}
\end{figure}

If the statistics are increased to 1.5*\nova + 2*T2K, as seen in 
Fig.~\ref{1.5nova2t2kdcp-sysNH}
then most of the
wrong hierarchy allowed region is ruled out as already noted in section 3.
For the true hierarchy, the allowed region is centered around true $\dcp$ 
and is mostly in the correct half-plane. For the
CP conserving case $\dcp = 0$ ($\dcp = \pm 180^\circ)$, there is a 
small additional allowed region around $\dcp = \pm 180^\circ$ ($\dcp = 0$) 
but for which $\chi^2$ is higher. If we limit our attention to
the regions around $\chi^2_{min}$, then 1.5*\nova + 2*T2K can 
measure $\dcp$ with an accuracy of 
$\pm 40^\circ$ for true $\dcp = 0$ and $\pm 60^\circ$ for 
true $\dcp = \pm 90^\circ$. 

It is curious that the CP conserving values of $\dcp$ 
can be measured with better accuracy than large CP violating values.
However, this point can be understood very simply in terms of 
Eq.~(\ref{pmue}). $\dcp$ occurs in this equation as 
$\cos(\hat{\Delta}+\dcp)$. Any experiment is designed such 
that the flux peaks at the energy where $\hat{\Delta} \approx 90^\circ$.
Thus the $\dcp$ term is approximately $-\sin \dcp$. The slope of 
$\sin x$ is large at $x \approx 0 \ {\rm or} \ 180^\circ$ and 
is very small at $x \approx \pm 90^\circ$. Therefore the uncertainty
in $\dcp$ is small near $0$ or $180^\circ$ and is large when
$\dcp$ is close to $\pm 90^\circ$.

Thus we are led to the following important conclusion:
1.5*\nova + 2*T2K can essentially determine the hierarchy
and also give an allowed region of $\dcp$ centered around its
true value.
Doubling of statistics will not lead to too much
improvement in the allowed range of $\dcp$. Further strategies
are needed to measure $\dcp$ to a good accuracy. 

A recent paper \cite{incremental} envisaged some future very long 
baseline superbeam experiments. 
They found that the early data from these will determine hierarchy, and 
additional data is needed to measure $\dcp$. 
We find that in the current scenario also, 
these considerations hold true.

\section{Summary}

In this paper we explored the hierarchy - $\dcp$ degeneracy 
of $\pmue$ of medium long baseline experiments. This degeneracy
severely limits the ability of any single experiment to determine
these quantities.  The observed moderately large value of $\ty$ is certainly 
a very good news for the upcoming NO$\nu$A, as it will lie in the
region where \nova has appreciable reach for hierarchy determination 
if the value of $\dcp$ happens to be favourable.
We define the concept of favourable half-plane of $\dcp$ and show that
the LHP(UHP) is the favourable(unfavourable) half-plane for NH
and vice-verse for IH. We also show that \nova by itself can determine
the hierarchy if $\dcp$ is in the favourable half-plane and 
$\sin^2 2 \ty \geq 0.12$. When $\dcp$ is in the unfavourable half-plane,
the data from \nova and T2K beautifully complement each other to rule 
out the wrong hierarchy.
We explore the underlying physics in detail and deduce the statistics 
needed for hierarchy determination. 
Given the current best fit of $\sin^2 2 \theta_{13} \simeq 0.1$,
the combined data from \nova and T2K can {\it essentially} resolve 
mass hierarchy for the entire $\dcp$ range if the statistics
for \nova and T2K are boosted by factors 1.5 and 2 respectively.
A baseline of $\sim$ 130 km will not be a bonus, over and above T2K, unless
supplemented by huge statistics.\\

In the last section we estimate the $\dcp$ reach of \nova and T2K. 
We demonstrate  
that without knowing the hierarchy, measuring $\dcp$ would be impossible.
With 1.5*\nova + 2*T2K, the allowed region of $\dcp$ is centered around
its true value and is mostly in the correct half-plane. Here also, a
short baseline of $\sim$ 130 km will not provide better information than T2K
with the same statistics.

\begin{acknowledgments}
 We thank Thomas Schwetz-Mangold for a comment on an earlier version, which 
led to an improvement of the manuscript. 
\end{acknowledgments}

\bibliographystyle{apsrev}
\bibliography{neutosc1}

\end{document}